\documentclass[prx,twocolumn,english,floatfix]{revtex4-2}

\usepackage{amsmath}
\usepackage{amssymb}
\usepackage{graphicx}
\usepackage{amsfonts}
\usepackage{physics}
\usepackage{comment}
\usepackage{float}
\usepackage{natbib}
\usepackage{color}
\usepackage{tikz}

\newcommand\checkmarkbis[1][]{
  \tikz[scale=0.4,#1]{\fill(0,.35) -- (.25,0) -- (1,.7) -- (.25,.15) -- cycle;}
}
\newcommand\crossmark[1][]{
  \tikz[scale=0.4,#1]{
    \fill(0,0)--(0.1,0) .. controls (0.5,0.4) .. (1,0.7)--(0.9,0.7) ..  controls (0.5,0.5) ..(0,0.1) --cycle;
    \fill(1,0.1)--(0.9,0.1) .. controls (0.5,0.3) .. (0,0.7)--(0.1,0.7) .. controls (0.5,0.4) ..(1,0.2) --cycle;
  }
}

\begin{document}

\title{Kinetic energy equipartion: a tool to characterize quantum thermalization}
\author{Carlos F. Destefani}
\altaffiliation[email: ]{carlos.destefani@uab.es}
\author{Xavier Oriols}
\altaffiliation[email: ]{xavier.oriols@uab.es}
\affiliation{Department of Electronic Engineering, Universitat Aut\`onoma de Barcelona, 08193 Bellaterra, Barcelona, Spain}

\begin{abstract}

Weak values have gradually been transitioning from a theoretical curiosity to a practical novel tool to characterize quantum systems in the laboratory.   When built by post-selecting the position, weak values of the momentum are linked to the so-called hidden varaiables of the Bohmian and Stochastic quantum mechanics. According to both theories,  the Orthodox kinetic energy has, in fact, two hidden-variable components: one linked to the \textit{current} (or Bohmian) velocity, and another linked to the \textit{osmotic} velocity (or quantum potential), and which are respectively identified with phase and amplitude of the wavefunction. Inspired by such alternative formulations, we address what happens to each of these two velocity components when the Orthodox kinetic energy thermalizes in closed systems, and how the pertinent weak values yield experimental information about them. We show that, after thermalization, the expectation values of both the (squared) current and osmotic velocities approach the same stationary value, that is, each of the Bohmian kinetic and quantum potential energies approaches half of the Orthodox kinetic energy. Such a `kinetic energy equipartition' is a novel signature of quantum thermalization that can empirically be tested in the laboratory, following a well-defined operational protocol as given by the expectation values of (squared) real and imaginary parts of the local-in-position weak value of the momentum, which are respectively related to the current and osmotic velocities. Thus, the kinetic energy equipartion presented here is independent on any ontological status given to these hidden variables, and it could be used as a novel element to characterize quantum thermalization in the laboratory, beyond the traditional use of expectation values linked to Hermitian operators. Numerical results for the nonequilibrium dynamics of a few-particle harmonic trap under random disorder are presented as illustration. And the advantages in using the center-of-mass frame of reference for dealing with systems with many indistinguishable particles are also discussed.  

\end{abstract}

\maketitle

\section{Introduction}
\label{intro}

A renewed interest in statistical mechanics of closed quantum systems has arisen \cite{reviewcoldatoms,reviewclosesystem,quantumsimulation,laser,reimann,therm_rigol,annualrev,ETHreview_rigol,deutsch_review,exp_review,exp_review_icfo,palencia_review,quenched_RMP,Reimann_2015,equilibration} as a consequence of the successful experimental ability to isolate and manipulate bosonic \cite{1DBosegases1,1DBosegases2,1DBosegases3,bosonicexpansion,dissipativeBEC} and fermionic \cite{fermionic,fermilattice,fermioptical,dipolefermi,fermionic_transport} many-body systems built on ultra-cold atomic gases subjected to optical lattices. The main question to be addressed is when an initial nonequilibrium state thermalizes and, if so, in which conditions. The Eigenstate Thermalization Hypothesis (ETH) \cite{ETH_origin,deutsch_91}, which has become a cornerstone in the study of quantum thermalization, claims that all relevant energy eigenstates of a given Hamiltonian, in the description of a quantum state, are thermal in the sense that they are similar to an equilibrium state as long as one deals with macroscopic observables. In recent years a large amount of numerical experiments has successfully tested ETH by directly diagonalizing in physical space some sort of short-range many-body lattice Hamiltonian, like Fermi- or Bose-Hubbard \cite{1DBosegases2,fermilattice,fermioptical,traject_measure,fermionic_off_rigol,lea_rigol_example} and XXZ- or XYZ-Heisenberg \cite{therm_rigol,equilibration,1DBosegases3,exp_rigol,nonintegrable_gogolin,rigol_ref16,Rigol_XXZeasyplane}, in the search of chaotic signatures in the statistics of their spectra, as in general induced by local impurities, without the need to explicitly evolve the initial nonequilibrium state. The true time-evolution of such a state is not as widespread due to the inherently huge configuration space involved.

Our understanding about quantum thermalization has largely been based in terms of expectation values of observables linked to Hermitian operators. It is well-known that other explanations of quantum phenomena allow one to discuss properties not directly linked to Hermitian operators. Such alternative explanations are in general labeled as hidden-variable theories. The Bohmian theory, formulated by de Broglie in 1927 \cite{debroglie} and further developed by Bohm in 1952 \cite{bohm}, is the most well-known example of a quantum theory with additional microscopic variables: particles have always well-defined positions that conform trajectories. Another example is the Stochastic quantum mechanics, proposed by Nelson in 1966 \cite{nelson}; although it also assumes particles with well-defined trajectories, these are unknown and only their statistical behavior is handled.  
   
These two hidden-variable theories assume that the Orthodox kinetic energy in fact has two components: in the Bohmian theory, it is computed as the square of the so-called Bohmian velocity plus a quantum potential, while in the Stochastic quantum mechanics, it is given by the square of a mean current velocity plus the square of a so-called osmotic velocity. The central question in this paper is whether these two hidden-variable components of the kinetic energy also thermalize when the Orthodox kinetic energy thermalizes.  We show that, in fact, one can characterize the thermalization time $t_{eq}$ as the time when the expectation value of the (squared) current and osmotic velocities become equal or, similarly, when the expectation values of the Bohmian kinetic and quantum potential energies become the same, with each being equal to half of the Orthodox kinetic energy. Such a \textit{kinetic energy equipartition} is the hidden-variable signature of quantum thermalization. 

Historically, hidden variables were thought to be hidden in the sense that they were not measurable in the laboratory, so that they were useful only as a complementary visualization tool of quantum dynamics. However, nowadays, it is well-known that one may have   empirical access to them through the so-called weak values, which have a well-defined operational protocol, independent on any ontology assigned to such hidden variables. Since  the original single-particle proposal \cite{weakvalue1988} weak values have been attracting a lot of theoretical \cite{weakvalue2021,wiseman2007grounding,durr2009weak,Marian16,velocity} and experimental \cite{kocsis2011observing,hariri2019experimental,ramos2020measurement} interests in distinct research fields, and a many-particle generalization   has been presented by the authors   elsewhere \cite{ourPRA}. One should clarify though that the fact that one may have experimental access to hidden variables does not directly imply that they are ontologically real. Each quantum theory defines its own ontological elements. For example, the Bohmian velocity is part of the ontology of the Bohmian mechanics (as the field which drives the particle trajectories), but not of the Orthodox ontology. And, in fact, in the Stochastic quantum mechanics the Bohmian velocity itself is not a direct ontological element, but part of a mean current velocity. The discussion of such ontological status is far from the scope of our paper. Our main focus is to emphasize   that the existence of the local-in-position weak values of the momentum opens a new unexplored link between theoretical predictions  and empirical data   that allows a novel characterization of quantum systems.  One could use other types of weak values, but working with the local-in-position weak values of the momentum allows us to re-use all the mathematical machinery (without choosing their ontology) of the Bohmian and Stochastic quantum mechanics.   And for that, in particular, we focus here on how the expectation values of the (squared) real and imaginary parts of such new empirical data allow one to characterize the process of quantum thermalization of a fermionic few-body closed system, defined by an harmonic trap under random disorder. At the end of the paper, we reformulate our proposal in the center-of-mass frame, so that our findings are also extendable for larger systems with many indistinguishable particles. 

The paper is organized as follows. Section \ref{theory} presents a brief summary of needed theoretical background. Section \ref{equipartition} addresses the Ortodox kinetic energy equipartition among its two hidden-variable components. Section \ref{results} presents the theoretical model and the numerical results for a particular few-body harmonic trap under random disorder, within three distinct scenarios for the dynamics; the center-of-mass frame is then considered for approaching larger systems. In Section \ref{conclusion} we conclude.

\section{Theoretical background} 
\label{theory}

This section provides a summary of the needed theoretical background: subsection \ref{WF} presents the weak values from the polar form of the many-body wave function, as well as the main equations of both Bohmian quantum mechanics and stochastic quantum mechanics; subsection \ref{EXin} develops the non-hermitian expectation values derived from such weak values.  To simplify the notation,   atomic units are employed throughout the text, which assumes $N$ nonrelativistic spinless particles each in a $1$D physical space, so that the position in configuration space is $\mathbf x = (x_1,...,x_N)$. 

\subsection{Weak values post-selected in position}
\label{WF}

A simple path to define the (complex) local-in-position weak value of the momentum for particle $j$ at position $\mathbf x$, $p_{W,j}(\mathbf{x},t)$, linked to the hermitian operator for the momentum, $\hat{p}_j = -i \; \partial / \partial x_j$, comes after inserting the identity $\int d\mathbf{x} |\mathbf{x}\rangle\langle \mathbf{x}|$ into the expectation value $\langle \Psi(t)|\hat{p}_j|\Psi(t)\rangle$,
\begin{equation}
\langle \hat p_j \rangle=\langle \Psi(t)|\hat{p}_j|\Psi(t)\rangle=\int d\mathbf{x} |\Psi (\mathbf{x},t)|^2\frac{\langle \mathbf{x}|\hat p_j|\Psi(t)\rangle}{\langle \mathbf{x}|\Psi(t)\rangle}.
\label{dPSI}
\end{equation}
From the polar form of the wave function, $\Psi (\mathbf{x},t) = R(\mathbf{x},t) \text{exp}(i S(\mathbf{x},t))$ with $|\Psi (\mathbf{x},t)|^2 = R^2 (\mathbf{x},t)$, the weak value as stated in \eqref{dPSI} decomposes like
\begin{eqnarray}
p_{W,j}(\mathbf{x},t) &\equiv& \frac{\langle \mathbf{x}|\hat p_j|\Psi(t)\rangle}{\langle \mathbf{x}|\Psi(t)\rangle} =\frac{\partial S(\mathbf{x},t)}{\partial x_j}-i \; \frac{1}{R(\mathbf{x},t)}\frac{\partial R(\mathbf{x},t)}{\partial x_j}, \nonumber\\
&=&  v_j(\mathbf{x},t) - i \; u_j(\mathbf{x},t), \label{WVp} 
\end{eqnarray}
where the current $v_j(\mathbf{x},t)$ and osmotic $u_j(\mathbf{x},t)$ velocities for particle $j$ are identified in \eqref{WVp} as
\begin{eqnarray}
v_j (\mathbf{x},t) &=& \frac{\partial S(\mathbf{x},t)}{\partial x_j}                                           = \text{Im} \left[\frac{1}{\Psi(\mathbf{x},t)} \frac{\partial\Psi (\mathbf{x},t)}{\partial x_j} \right], \label {DEFvj} \\
u_j (\mathbf{x},t) &=& \frac{1}{R(\mathbf{x},t)}\frac{\partial R(\mathbf{x},t)}{\partial x_j} = \text{Re}\left[ \frac{1}{\Psi (\mathbf{x},t)}\frac{\partial \Psi(\mathbf{x},t)}{\partial x_j}\right]. \label{DEFuj}
\end{eqnarray}
Notice that $v_j (\mathbf{x},t)$ and $u_j (\mathbf{x},t)$ depend only, respectively, on the phase and on the amplitude of $\Psi (\mathbf{x},t)$. The strategy above in fact can be used for any operator; for the bilinear momentum, $\hat p_j \hat p_l = -\partial / \partial x_j \partial / \partial x_l$, one has
\begin{equation}
\langle \hat p_j \hat p_l \rangle=\langle \Psi(t)| \hat p_j \hat p_l|\Psi(t)\rangle=\int d\mathbf{x} |\Psi (\mathbf{x},t)|^2\frac{\langle \mathbf{x}|\hat p_j \hat p_l|\Psi(t)\rangle}{\langle \mathbf{x}|\Psi(t)\rangle},
\label{dPSI2}
\end{equation}
where the bilinear weak value $p_{W,jl}(\mathbf{x},t)$ is
\begin{eqnarray}
&&p_{W,jl}(\mathbf{x},t) \equiv \frac{\langle \mathbf{x}|\hat p_j \hat p_l|\Psi(t)\rangle}{\langle \mathbf{x}|\Psi(t)\rangle} \nonumber \\
&&=v_j(\mathbf{x},t)  v_l(\mathbf{x},t) - u_j(\mathbf{x},t)  u_l(\mathbf{x},t) - \frac{\partial  u_l(\mathbf{x},t)}{\partial x_j} \label{WVpp} \\
&&- i \; \left( u_j(\mathbf{x},t)  v_l(\mathbf{x},t) + v_j(\mathbf{x},t)  u_l(\mathbf{x},t) + \frac{\partial  v_l(\mathbf{x},t)}{\partial x_j} \right), \nonumber
\end{eqnarray}
which, for $j=l$, defines the local-in-position weak value of (twice) the kinetic energy for particle $j$, $K_{W,j}(\mathbf{x},t)$,
\begin{eqnarray}
&&2 K_{W,j}(\mathbf{x},t) \equiv \frac{\langle \mathbf{x}|\hat p_j^2|\Psi(t)\rangle}{\langle \mathbf{x}|\Psi(t)\rangle} = v^2_j(\mathbf{x},t) - u^2_j(\mathbf{x},t)\nonumber \\ 
&&  - \frac{\partial  u_j(\mathbf{x},t)}{\partial x_j}  - i \; \left( 2 u_j(\mathbf{x},t)  v_j(\mathbf{x},t)  + \frac{\partial  v_j(\mathbf{x},t)}{\partial x_j} \right).
\label{WVp2}
\end{eqnarray}
Subsecion \ref{EXin} will show that the imaginary part does not contribute for ensemble values of \eqref{WVp}, \eqref{WVpp}, or \eqref{WVp2}. One could also define from \eqref{WVp} and \eqref{WVpp} the (weak) correlation of the momentum post-selected in position, $C_{p_{W,j}; p_{W,l}} \equiv \langle p_{W,jl}(\mathbf{x},t) \rangle - \langle p_{W,j}(\mathbf{x},t) \rangle \langle p_{W,l}(\mathbf{x},t) \rangle$, as
\begin{equation}
C_{p_{W,j};  p_{W,l}} = - \langle \frac{\partial  u_l(\mathbf{x},t)}{\partial x_j} \rangle - i \; \langle \frac{\partial  v_l(\mathbf{x},t)}{\partial x_j} \rangle,
\label{twoWV}
\end{equation}
which could be used in situations where separated entanglements in phase and amplitude of $\Psi(\mathbf{x},t)$ were accessible \cite{WVentang}, since the real (imaginary) part of $C_{p_{W,j};  p_{W,l}}$ depends only on the amplitude (phase) of $\Psi(\mathbf{x},t)$. Some given properties among current \eqref{DEFvj} and osmotic \eqref{DEFuj} velocities will prove important in our derivation. For that,  one can use  some basic elements of both Bohmian and Stochastic quantum mechanics. 

\subsubsection{Bohmian quantum mechanics}
\label{BQM}

The Bohmian theory assumes that particles follow well-defined trajectories guided by $\Psi (\mathbf{x},t)$, solution of the Schroedinger equation
\begin{equation}
 i\frac{\partial\Psi (\mathbf{x},t)}{ \partial t} =  H (\mathbf{x})\Psi (\mathbf{x},t), 
\label{scho}
\end{equation}
whose Hamiltonian is
\begin{equation}
H (\mathbf{x})=-\frac{1}{2} \sum_{j=1}^N \frac{\partial^2}{\partial x_j^2}+V (\mathbf{x}) = K (\mathbf{x}) + V (\mathbf{x}),
\label{hamiltonian}
\end{equation}
with $K (\mathbf{x}) = \sum_{j=1}^N K_j (x_j)$, and $K_j (x_j) \equiv -\partial^2 / 2 \partial x_j^2$ the kinetic energy of each of the $N$ particles, whose interaction is given by the potential energy $V(\mathbf{x})$. From the polar form of $\Psi (\mathbf{x},t)$, equation \eqref{scho} gets rewritten as two coupled equations. On one hand, the imaginary part yields the continuity equation,
\begin{equation}
\frac{\partial  R^2(\mathbf{x},t)}{\partial t} + \sum_{j=1}^N R^2(\mathbf{x},t) \frac{\partial S(\mathbf{x},t)}{\partial x_j} = 0 ,
\label{continuity}
\end{equation} 
in which from \eqref{DEFvj} one identifies the $j$-component of the current density as $J_j (\mathbf{x},t) \equiv  R^2(\mathbf{x},t) v_j(\mathbf{x},t)$. On the other hand, the real part yields the quantum Hamilton-Jacobi equation,
\begin{equation}
\frac{\partial S(\mathbf{x},t)}{\partial t} + K_B(\mathbf{x},t) + Q_B(\mathbf{x},t)+V(\mathbf{x})=0,
\label{HamJac1}
\end{equation}
with $K_B(\mathbf{x},t)=\sum_{j=1}^N K_{B,j}(\mathbf{x},t)$ and $Q_B(\mathbf{x},t)=\sum_{j=1}^N Q_{B,j}(\mathbf{x},t)$, and where the $j$-component of the Bohmian kinetic $K_{B,j}(\mathbf{x},t)$ and quantum potential $Q_{B,j}(\mathbf{x},t)$ energies are  
\begin{eqnarray}
K_{B,j}(\mathbf{x},t) &=&  \frac{1}{2} \left( \frac{\partial S(\mathbf{x},t)}{\partial x_j} \right) ^2, \label{KB0} \\
Q_{B,j}(\mathbf{x},t) &=& -\frac{1}{2 R(\mathbf{x},t)}\frac{\partial^2 R(\mathbf{x},t)}{\partial x^2_j}. \label{QB0}
\end{eqnarray}
The Bohmian trajectories, thanks to the current density $J_j (\mathbf{x},t)$ in \eqref{continuity}, are obtained from the integration solely of $v_j (\mathbf{x},t)$ ($u_j (\mathbf{x},t)$ plays no role); so, along the text $v_j (\mathbf{x},t)$ is interchangeably identified either as Bohmian or current velocity. In Bohmian theory, while the kinetic energy is determined by the phase of the wavefunction, the quantum potential, determined by its amplitude, has indeed the status of a potential energy since, from a time derivative in \eqref{DEFvj} and by using \eqref{HamJac1}, one gets the set of Newton-like equations,  
\begin{equation}
\frac{d ^2 x_j(t)}{d t^2} = -\left[ \frac{\partial}{\partial x_j} \left(  V(\mathbf{x},t) + Q_B(\mathbf{x},t)  \right) \right]_{\mathbf{x}=\mathbf{x(t)}},
\label{newton}
\end{equation}
which is another way for getting Bohmian trajectories.

\subsubsection{Stochastic quantum mechanics}
\label{SQM}

The Stochastic quantum mechanics may be understood as an attempt to give a kinematic interpretation also to the quantum potential, although such a potential is not explicit in its original derivation. Notice indeed that, from \eqref{DEFvj}-\eqref{DEFuj}, equations \eqref{KB0}-\eqref{QB0} can be cast into
\begin{eqnarray}
K_{B,j}(\mathbf{x},t) &=&  \frac{1}{2} v_j^2(\mathbf{x},t) , \label{KBj_vj} \\
Q_{B,j}(\mathbf{x},t) &=& -\frac{1}{2} u_j^2(\mathbf{x},t) - \frac{1}{2}\frac{\partial u_j(\mathbf{x},t)}{\partial x_j}, \label{QBj_uj} 
\end{eqnarray}
so that $K_{B,j}(\mathbf{x},t)$ and $Q_{B,j}(\mathbf{x},t)$, respectively, depend only on Bohmian and osmotic velocities. Notice that $K_{B,j}(\mathbf{x},t)$ and $Q_{B,j}(\mathbf{x},t)$ are not weak values computed from the hermitian operator of the kinetic energy, as it happens with $K_{W,j}(\mathbf{x},t)$ in \eqref{WVp2}. One should keep in mind that, according to this stochastic theory, $v_j(\mathbf{x},t)+u_j(\mathbf{x},t)$ has the meaning of a mean velocity, while the true velocity is a random velocity around it. Such a theory is defined in terms of a stochastic diffusion process in the configuration space, which requires that the probability $R^2 (\mathbf{x},t)$ satisfies both forward ($+$) and backward ($-$) Fokker-Plank equations for a parameter $\nu$, 
\begin{eqnarray}
\frac {\partial R^2 (\mathbf{x},t)}{\partial t}&=&-\sum_j \frac{\partial }{\partial x_j} \left( (v_j (\mathbf{x},t) \pm u_j (\mathbf{x},t) )R^2 (\mathbf{x},t) \right) \nonumber \\
                                                                         &\pm& \nu \sum_j \frac{\partial^2 R^2 (\mathbf{x},t)}{\partial x_j^2},
\label{fokker}
\end{eqnarray}
whose sum yields the continuity equation in \eqref{continuity}, irrespective of $u_j (\mathbf{x},t)$ and $\nu$, and whose difference yields 
\begin{equation}
\nu \frac {\partial^2 R^2 (\mathbf{x},t)} {\partial x_j^2} = \frac {\partial \left( u_j (\mathbf{x},t) R^2 (\mathbf{x},t) \right)} { \partial x_j}, 
\label{diffcurr}
\end{equation}
which is satisfied by the $u_j (\mathbf{x},t)$ in \eqref{DEFuj} and with $\nu = 1/2$. 
In other words, the Fokker-Planck equations in \eqref{fokker}, with \eqref{DEFuj} and $\nu=1/2$, pictures from \eqref{diffcurr} that the osmotic current $u_j (\mathbf{x},t)R^2 (\mathbf{x},t) $ is balanced by some diffusion current $\nu \partial R^2 (\mathbf{x},t) / \partial x_j$, so that the continuity equation in \eqref{continuity} remains valid and determined solely by the current velocity $v_j (\mathbf{x},t)$. As such, the kinematic interpretation of the quantum potential, implicit in the derivation of the Stochastic quantum mechanics, empirically reproduces both Bohmian and Orthodox quantum mechanics. 

\subsection{Expectation values from weak values}
\label{EXin}

From \eqref{dPSI}-\eqref{WVp} the expectation value for the momentum weak value post-selected in position, $\langle p_{W,j} \rangle$, is
\begin{equation}
\langle \hat p_j \rangle  =  \int  d\mathbf{x}  \; p_{W,j}(\mathbf{x},t) |\Psi (\mathbf{x},t) |^2 \equiv \langle p_{W,j} \rangle.
\label{EVp}
\end{equation}
For this particular case one can directly compute the expectation values of each of its real and imaginary parts, yielding from \eqref{dPSI}-\eqref{WVp}
\begin{eqnarray}
\langle v_j \rangle  &=&  \int  d\mathbf{x}  \; \text{Re}\left[p_{W,j}(\mathbf{x},t)\right] |\Psi (\mathbf{x},t) |^2 = \langle \hat p_j \rangle , \label{int_v} \\
\langle u_j \rangle  &=&  \int  d\mathbf{x}  \; \text{Im}\left[p_{W,j}(\mathbf{x},t)\right] |\Psi (\mathbf{x},t) |^2=0, \label{int_u}
\end{eqnarray}
so that,  while $ v_j (\mathbf{x},t) \neq p_{W,j}(\mathbf{x},t)$,  $\langle v_j \rangle = \langle p_{W,j} \rangle$ and the osmotic velocity has no role in the expectation values of neither $\hat p_j$ nor $p_{W,j}(\mathbf{x},t)$ (for a wave function vanishing at the system borders). Strictly speaking, $v_j(\mathbf{x},t)$ and $u_j(\mathbf{x},t)$ are not weak values themselves but they are, respectively, post-processed real and imaginary parts of the weak value $p_{W,j}(\mathbf{x},t)$. In other words, while $p_{W,j}(\mathbf{x},t)$ is linked to the hermitian operator $\hat p_j$, no hermitian operators can be linked to $v_j(\mathbf{x},t)$ and $u_j(\mathbf{x},t)$. The connection among $p_{W,j}(\mathbf{x},t)$ and $v_j(\mathbf{x},t)$ is approached elsewhere \cite{review2012,review2014,hydroBM,hydro1,hydro2,hydro3,Oriols12,ontologies}, while more attention has recently been paid to the meaning of $u_j(\mathbf{x},t)$ \cite{WVtomo,WVim1,WVstrong,QPinternal,QPhydro,QPthermo,QPsanz,BFmom,WVFP}. We distinguish in the paper three types of expectation values: (i) $\langle \hat a \rangle$ with ``hat''  for the operator $\hat a$; (ii) $\langle a_W \rangle $ with subindex ``W'' for the weak value $a_W$; (iii) $\langle a \rangle $ for the value $a$ obtained by post-processing the weak value.   

The expectation value of the bilinear weak value, $\langle p_{W,jl} \rangle$, is from \eqref{dPSI2}-\eqref{WVpp}
\begin{eqnarray}
\langle \hat p_j \hat p_l \rangle  &=&  \int  d\mathbf{x}  \; p_{W,jl}(\mathbf{x},t) |\Psi (\mathbf{x},t) |^2  \rangle, \nonumber \\
 &=& \langle v_j v_l \rangle + \langle u_j u_l \rangle \equiv \langle p_{W,jl} \rangle ,
\label{EVp2}
\end{eqnarray}
where we have used 
\begin{eqnarray}
\left \langle \frac{\partial u_l} {\partial x_j} \right \rangle  &=& -2 \langle u_l u_j\rangle ,  \label{int_du} \\
\left \langle \frac{\partial v_l} {\partial x_j} \right \rangle  &=& -\langle u_l  v_j \rangle  -\langle v_l  u_j \rangle  \label{int_dv}; 
\end{eqnarray}
for an antisymmetric $\Psi (\mathbf{x},t)$ one has $\langle u_l  v_j \rangle  = \langle v_l  u_j \rangle$. Once more, the weak value $p_{W,jl}(\mathbf{x},t)$ is linked to $\hat p_j \hat p_l$, but $v_j(\mathbf{x},t) v_l(\mathbf{x},t)$ and $u_j(\mathbf{x},t) u_l(\mathbf{x},t)$ are just post-processed data obtained from  the real and imaginary parts of   the weak values $p_{W,j}(\mathbf{x},t)$ and $p_{W,l}(\mathbf{x},t)$. Interestingly, from \eqref{int_du} the expectation values of the Bohmian kinetic $\langle K_{B,j} \rangle $ and quantum potential $\langle Q_{B,j} \rangle $ energies in \eqref{KBj_vj} and \eqref{QBj_uj} become 
\begin{eqnarray}
\langle K_{B,j} \rangle  &=&  \frac{1}{2} \langle v_j^2\rangle , \label{BKE} \\
\langle Q_{B,j} \rangle &=&  \frac{1}{2} \langle u_j^2\rangle , \label{BQP} \\
\langle\hat  K_j \rangle   &=&  \frac{1}{2}  (\langle v^2_j \rangle + \langle u^2_j \rangle), \label{OKE}
\end{eqnarray}
where the last equation is the expectation value of the Orthodox kinetic energy, $\langle \hat K_j \rangle =\langle \hat p^2_j \rangle  / 2$, as promptly obtained from \eqref{EVp2}. Neither $\langle v_j^2\rangle $ nor $\langle u_j^2\rangle $ are expectation values of a weak value but instead, respectively, of the (squared) real and imaginary parts of the weak value $p_{W,j}(\mathbf{x},t)$. Notice that by integrating the weak values of the kinetic energy in \eqref{WVp2} with the use of \eqref{int_du}-\eqref{int_dv}, and comparing the result with \eqref{OKE}, one obtains  
\begin{equation}
\langle \hat K_j \rangle  =  \int  d\mathbf{x} \; K_{W,j}(\mathbf{x},t) |\Psi (\mathbf{x},t) |^2 =  \langle K_{W,j} \rangle.
\label{EVWVp2}
\end{equation}

Similar to the (weak) correlation in \eqref{twoWV}, one can define the correlations $C_{\hat a,\hat b} \equiv \langle \hat a \hat b \rangle - \langle \hat a \rangle  \langle \hat  b \rangle$ between operators and $C_{a,b} \equiv \langle a b \rangle - \langle a \rangle  \langle  b \rangle $ between quantities post-processed from weak values. For the momentum operator, from \eqref{EVp2} and \eqref{int_v} one obtains
\begin{eqnarray}
&&C_{\hat p_j,\hat p_l}     =  \langle v_j v_l \rangle + \langle u_j u_l \rangle - \langle v_j \rangle  \langle  v_l \rangle =  C_{v_j,v_l} +  C_{u_j,u_l}, \label{twoC} \\
&&C_{p_{W,j};  p_{W,l}} = 2 \langle u_j u_l\rangle  + i \left(  \langle u_j  v_l \rangle  + \langle v_j  u_l \rangle \right),\label{twoCweak}
\end{eqnarray}
where \eqref{int_u} is used in \eqref{twoC}, while \eqref{twoCweak} results from applying \eqref{int_du}-\eqref{int_dv} in \eqref{twoWV}. So, while the Bohmian velocity fully determines $\langle \hat p_j \rangle$ in \eqref{int_v}, the osmotic velocity, although satisfying \eqref{int_u}, induces a deviation between the quantum correlations of the momentum with respect to the curent velocity in \eqref{twoC}. Notice that from \eqref{twoC}-\eqref{twoCweak} one has $\text{Re} [C_{p_{W,j};  p_{W,l}}] = 2 C_{u_j,u_l}$, while one should have $\text{Im} [C_{p_{W,j};  p_{W,l}}] \approx 0$.

\section{Equipartion at thermalization of the hidden-variable components}
\label{equipartition}
 
Since the expectation values $\langle K_{B,j} \rangle$ and $\langle Q_{B,j} \rangle$ cannot be computed from hermitian operators, the typical argumentation of ETH to discuss thermalization of observables cannot directly be applied to these quantities, although its very same spirit can still be used. For that, let us first summarize the standard interpretation of the thermalization of expectation values. 

From an initial nonequilibrium state $|\Psi(0)\rangle$, whose unitary evolution is dictated by $|\Psi(t)\rangle=\sum_n c_n e^{-i \omega_n t} |n\rangle$, being $|n \rangle$ an energy eigenstate with eigenvalue $\omega_n$, and $c_n=\langle n|\Psi(0)\rangle$ defined by initial conditions, the expectation value of some hermitian operator $\hat a$ reads
\begin{equation}
\langle \hat a \rangle = \sum_n \rho_{n,n} (0) a_{n,n}+\sum_{n,m\neq n} \rho_{n,m} (t) a_{m,n} ,
\label{rho}
\end{equation}
with $a_{m,n}=\langle m |\hat a |n \rangle$, and $\rho_{n,m}(t)=c_m^* c_n e^{i(\omega_m-\omega_n)t}$ the density matrix, composed by diagonal time-independent and nondiagonal time-dependent terms; while the former can never be neglected (unless is zero by construction), the latter needs to be negligible after some given time if one expects $\langle \hat a \rangle$ to thermalize. A system is said to equilibrate if, after some time $t_{eq}$ enough for full dephasing between different energy eigenstates, the nondiagonal terms cancel out so that \eqref{rho} can solely be computed from the diagonal terms, $\langle \hat a \rangle \approx \sum_n \rho_{n,n} (0) a_{n,n}$, for most times $t>t_{eq}$. The system is then said to thermalize when $\langle \hat a \rangle$ becomes roughly equal to the expectation value as computed from its microcanonical density matrix. The ETH states that such dephasing is more typical to nondegenerated and chaotic many-body scenarios, where the nondiagonals $a_{m,n}$ in \eqref{rho} become exponentially smaller than the diagonals $a_{n,n}$. In other words, a non-equilibrium state whose evolution involves a large number of eigenstates (not necessarily of particles) is required. Such a picture immediately applies to the thermalization of the orthodox kinetic $K(\mathbf{x})$ and potential $V(\mathbf{x})$ operators in \eqref{hamiltonian}. For the hidden-variable components of the Bohmian and Stochastic kinetic energies, however, one needs an extra step, which renders us the main result of our work in what follows.

The Bohmian velocity for particle $j$ is, from \eqref{DEFvj}, 
\begin{equation}
v_j (\mathbf{x},t) = \text{Im} \left[\frac{1}{\Psi(\mathbf{x},t)} \frac{\partial\Psi (\mathbf{x},t)}{\partial x_j} \right] = \frac{1}{2i}\left(\frac{\tilde \Psi_j}{\Psi}-\frac{\tilde \Psi^*_j}{\Psi^*}\right),
\label{vjtilde}
\end{equation}
where $\tilde \Psi_j \equiv \partial \Psi(\mathbf{x},t) / \partial x_j$ and $\Psi \equiv \Psi(\mathbf{x},t)$. The ensemble value of the product $\langle v_j v_l \rangle$ is
\begin{eqnarray}
\label{result2}
\langle v_j v_l \rangle &=&\int d\mathbf{x} |\Psi(\mathbf{x},t)|^2 \: v_j(\mathbf{x},t) v_l(\mathbf{x},t)\nonumber \\
                                 &=&-\frac{1}{4}\int d\mathbf{x} \frac{1}{|\Psi|^2} \left(\tilde \Psi_j\Psi^*-\tilde \Psi_j^* \Psi \right)\left(\tilde \Psi_l\Psi^*-\tilde \Psi_l^* \Psi \right)\nonumber\\
                                 &=&\frac{1}{2}\int d\mathbf{x}\tilde \Psi_j \tilde \Psi_l^* \label{KBfull} \\
                                &-&\frac{1}{4}\int d\mathbf{x} \frac{1}{|\Psi|^2} \left(\tilde \Psi_j \Psi^*\tilde \Psi_l \Psi^*+\tilde \Psi_j^* \Psi \tilde \Psi_l^* \Psi \right). \label{result}\nonumber
\end{eqnarray}
The first integral is just half of $\langle \hat p_j \hat p_l \rangle = \langle (\hat p_j \Psi (t))^* | \hat p_l \Psi (t) \rangle = \int d\mathbf{x}\tilde \Psi_j \tilde \Psi_l^*$, which then follows the same thermalization process discussed after \eqref{rho} for hermitian operators. The second integral is the sum of one component plus its complex conjugate so that \eqref{KBfull} remains indeed real; from the decomposition $\Psi(\mathbf x,t)=\sum_n c_n e^{-i \omega_n t} \psi_n(\mathbf{x})$ one can rewrite 
\begin{eqnarray}
&& \int d\mathbf{x} \frac{1}{|\Psi|^2} (\tilde \Psi_j \Psi^*\tilde \Psi_l \Psi^*)= \label{Kfull2} \\ 
&& \sum_{a,b,c,d} c_a c_b^* c_c c_d^* e^{-iw_{a,b,c,d}t} \int d\mathbf{x} \; \frac{1}{|\Psi |^2} \tilde \psi_{a,j} \psi_b^* \tilde \psi_{c,l} \psi_d^*, \nonumber 
\end{eqnarray}
where $\tilde \psi_{s,q} \equiv \partial \psi_s(\mathbf{x}) / \partial x_q$ and $\psi_{s} \equiv \psi_s(\mathbf{x}) $, for $s=a,b,c,d$, $q=j,l$, and $w_{a,b,c,d}=w_a-w_b+w_c-w_d$. Notice that contrarily to \eqref{rho} and \eqref{result2}, there is no time-independent term because the denominator $|\Psi|^2\equiv |\Psi(\mathbf{x},t)|^2$ will be time-dependent even when $w_{a,b,c,d}=0$. Due to the assumed chaotic nature of the eigenstates, one can infer no spatial correlation between $\psi_{a,j} (\mathbf{x})$, $\psi_b^* (\mathbf{x})$, $\tilde \psi_{c,l} (\mathbf{x})$, $\psi_d^* (\mathbf{x})$; in addition, $|\Psi(\mathbf{x},t)|^2$ introduces randomness in the configuration space as the system thermalizes in the physical space \cite{ourPRA}. With these in mind \eqref{Kfull2} simply yields a sum of random numbers around zero with vanishing contribution, so that the thermalized value of \eqref{KBfull} is
\begin{equation}
\langle v_j v_l \rangle \approx \frac{\langle \hat p_j \hat p_l \rangle}{2}.
\label{KBtermal}
\end{equation}
From exactly the same procedure, since the osmotic velocity for particle $j$ is, from \eqref{DEFuj}, 
\begin{equation}
u_j (\mathbf{x},t) = \text{Re} \left[\frac{1}{\Psi(\mathbf{x},t)} \frac{\partial\Psi (\mathbf{x},t)}{\partial x_j} \right] = \frac{1}{2}\left(\frac{\tilde \Psi_j }{\Psi }+\frac{\tilde \Psi^*_j}{\Psi^* }\right),
\label{ujtilde}
\end{equation}
the ensemble value of the product $\langle u_j u_l \rangle$ will be exactly the same as in \eqref{KBfull}, but with a positive sign in the second integral. Thus, applying the same reasoning as above, one finds the thermalized value
\begin{equation}
\langle u_j u_l \rangle \approx \frac{\langle \hat p_j \hat p_l \rangle}{2},
\label{QBtermal}
\end{equation}
so that \eqref{EVp2} remains satisfied after thermalization.

Applying \eqref{KBtermal} and \eqref{QBtermal} with $j=l$ into \eqref{BKE}-\eqref{OKE} yields
\begin{eqnarray}
\langle \hat K_j\rangle & = & \langle K_{B,j} \rangle + \langle Q_{B,j} \rangle, \label{KQsum} \\
\frac{\langle \hat K_j\rangle}{2} &\approx& \langle K_{B,j} \rangle \approx \langle Q_{B,j} \rangle. \label{KQhalf} 
\end{eqnarray}
While \eqref{KQsum} is already true in \eqref{BKE}-\eqref{OKE}, and so it is valid at any scenario, thermalized or not, the \textit{kinetic energy equipartition} in \eqref{KQhalf}, being only valid at times $t > t_{eq}$, presents the \textit{hidden-variable signature} of quantum thermalization: Bohmian kinetic and quantum potential energies become equal, with each being equal to half of the orthodox kinetic energy. Similarly, while $\langle \hat p_j^2 \rangle = \langle v_j^2 \rangle + \langle u_j^2 \rangle$ applies at any time, at $t>t_{eq}$ $\langle \hat p_j^2 \rangle/2 \approx \langle v_j^2 \rangle \approx \langle u_j^2 \rangle$ should also apply. That is, thermalization also implies that (squared) Bohmian and osmotic velocities become equal, with each being equal to half of the (squared) orthodox momentum; in other words, information from phase and amplitude of the wave function become similar after $t>t_{eq}$, which is a result of the initially localized wave function, in a nonequilibrium dynamics, spreading almost homogeneously through the whole configuration space after the onset of thermalization. In addition to \eqref{KQsum}-\eqref{KQhalf}, and particularly for a randomly disordered harmonic trap, one should also find that
\begin{eqnarray}
\langle \hat H\rangle &=& \langle \hat K \rangle + \langle \hat V \rangle, \label{vir1} \\
\frac{\langle \hat H\rangle}{2} & \approx & \langle \hat K \rangle \approx \langle \hat V \rangle. \label{vir2} 
\end{eqnarray}
Equation \eqref{vir1} expresses the conservation of total energy, from the Hamiltonian in \eqref{hamiltonian} in a unitary evolution, valid at any time. Equation \eqref{vir2} tells us that the orthodox virial theorem, as one reaches some steady state at $t>t_{eq}$, is expected to be restated; that is, potential and kinetic energies should become equal, with each being equal to half of the total energy. It is assumed  in \eqref{vir1}-\eqref{vir2} that $ \langle \hat V \rangle$ mostly comes from the confining potential. At last, from \eqref{KBtermal} and \eqref{QBtermal} into \eqref{twoC}-\eqref{twoCweak}, the thermalized correlations should satisfy
\begin{equation}
\frac{C_{\hat p_j,\hat p_l}}{2} \approx C_{v_j,v_l} \approx C_{u_j,u_l} \approx \frac{C_{p_{W,j}; p_{W,l}}}{2}.
\label{Cb}
\end{equation}

\section{Numerical results}
\label{results}

We first describe the non-equilibrium initial state and the Hamiltonian of our model; then we discuss our results according to three scenarios with distinct initial conditions. At the end we reformulate our model in the center of mass frame and address the respective results. 

\subsection{Initial state and trap Hamiltonian}
\label{HAM}

The initial $N$-electron non-equilibrium pure antisymmetric state is
\begin{equation}
\langle \mathbf{x}|\Psi(0) \rangle=\frac{1}{\mathcal{C}} \sum_{n=1}^{N!} \text{sign}(\vec p(n))\prod_{j=1}^{N} \psi_j(x_{p(n)_j},0), 
\label{initWF}
\end{equation}
with $\mathcal{C}$ a normalization constant and $\text{sign}(\vec p(n))$ the sign of the permutation $\vec p(n)=\{p(n)_1,..,p(n)_N\}$. Each initial Gaussian state in \eqref{initWF} is 
\begin{equation}
\psi_j(x,0)=\exp \left[ -\frac{(x-x_{0j})^{2}} { 2\sigma_{j}^{2}} \right] \exp \left[ ip_{0j}(x-x_{0j}) \right],
\label{initgauss} 
\end{equation}
with spatial dispersion $\sigma_{j}$, central position $x_{0j}$, and central velocity $p_{0j}$. Any nonzero $x_{0j}$ or $p_{0j}$ may activate the nonequilibrium dynamics.

The Hamiltonian $H(\mathbf{x})$ propagating the many-body wavefunction $\Psi (\mathbf{x},t)$ is given in \eqref{hamiltonian}, where the kinetic term $K (\mathbf{x})$ is already defined. The potential term $V (\mathbf{x})$, in our disordered harmonic trap, is defined by
\begin{equation}
V (\mathbf{x}) \equiv V_H (\mathbf{x}) + V_I (\mathbf{x}) + V_D (\mathbf{x}), 
\label{numV}
\end{equation}
where the harmonic potential $V_H (\mathbf{x})$ with frequency $\omega$ is 
\begin{equation}
V_H (\mathbf{x}) =  \frac{1}{2} \omega^{2} \sum_{j=1}^N  x_j^{2}, 
\label{numVH}
\end{equation}
and the electron-electron interaction potential $V_I (\mathbf{x})$ with smooth parameter $\alpha$ is  
\begin{equation}
V_I (\mathbf{x}) =  \frac{1}{2} \sum_{j=1}^N  \sum_{l \neq j}^N \frac{1}{\sqrt{(x_{j}-x_{l})^{2}+\alpha^{2}}}. 
\label{numVI}
\end{equation}
To ensure that the initial state in \eqref{initWF} is built as a superposition of a large (and `chaotic') number of eigenstates, we include the random disorder potential $V_D (\mathbf{x})$, 
\begin{equation}
V_D(\mathbf{x}) = \gamma_D \sum_{j=1}^N   \sum_{l=1}^{M}b_{l}\exp[-\frac{4(x_j-g_l)^{2}}{\sigma_D^{2}}] 
\label{numD},
\end{equation}
where $\gamma_D$ is its strength and $\sigma_D$ its spatial dispersion, with $g_{l}$ running through $M$ grid points; the set of random numbers $b_{l}$ satisfies $\langle b_{l} \rangle=0$ and $\langle b_{l}^{2} \rangle=1$, and the disorder potential is normalized so that the integral of $V^2_D (\mathbf{x})$ yields $\gamma_D^2$. Such a shape is typical of speckle potentials \cite{entropytoy,fermitrap,fermi2,virialEE,disorder_ong19,speckle,localspec} in fermionic traps. 

\subsection{Expectations values in three scenarios}
\label{EV3}

Our results initially focus on $N=2$. However our main conclusions, as the kinetic energy equipartition at thermalization, are valid for any $N$ as the discussion in the center of mass frame will latter settle. We focus on three scenarios, D1, D2, D3, which are distinct by the initial values of $x_{0j}$ and $p_{0j}$, as summarized in Table \ref{table}, where each scenario considers both no-disordered and disordered dynamics. Values of $x_{0j}$ in D1 and $p_{0j}$ in D2 are chosen identical as to yield the same turning points in both dynamics; the remaining simulation parameters are found in \cite{param}. The no-disorder cases employ smaller values of $x_{0j}$ and $p_{0j}$ to render the features more visible. 

The dynamics for each scenario D1, D2,  D3 is respectively shown in figures \ref{fig12}, \ref{fig34}, \ref{fig56}: top panels (a)-(d) show a few initial cycles with no disorder, bottom panels (e)-(h) show the full evolution with disorder. The structure of these three figures, which focus on the time evolution of some pertinent expectation values, is: panels (a), (e) show kinetic $\langle K \rangle$, harmonic $\langle V_H \rangle$, interaction $\langle V_I \rangle$, and total $\langle H \rangle = \langle K \rangle+\langle V_H \rangle+\langle V_I \rangle+\langle V_D \rangle$ energies, with disorder energy $\langle V_D \rangle \approx 0$ at any $t$ not shown; panels (b), (f) show Bohmian velocity $\langle v_j \rangle$, position $\langle x_j \rangle$, and osmotic velocity $\langle u_j \rangle$, with momentum $\langle p_j \rangle = \langle v_j \rangle$ not shown, and the label $j$ not indicated since it is redundant in our antisymmetrized model; panels (c), (g) repeat the orthodox kinetic energy $\langle K \rangle$, for its comparison with the Bohmian kinetic $\langle K_B \rangle$ and quantum potential $\langle Q_B \rangle$ energies; panels (d), (h) show the correlations for momentum $C_{\hat p_1,\hat p_2}$, Bohmian $C_{v_1,v_2}$ and osmotic $C_{u_1,u_2}$ velocities, and position $C_{\hat x_1,\hat x_2}$.
\begin {table}
\label{tab:table}
\begin{tabular}{c|cc|cc|cc}
\hline
\multicolumn{1}{c}{\textbf{Scenario}} & \multicolumn{2}{c}{\textbf{D1}} & \multicolumn{2}{c}{\textbf{D2}} & \multicolumn{2}{c}{\textbf{D3}} \\
\hline
\hline
 Disorder    & No  & Yes & No   & Yes & No  & Yes \\
\hline
\hline
  ($x_{01}$,$x_{02}$) &  (-2,2) & (-20,20) & (-2,2) & (-2,2) & (-2,2) & (-20,20)\\
\hline
  ($p_{01}$,$p_{02}$) &  (0,0) & (0,0) & (4,4) & (20,20) & (2,2) & (20,20)\\
\hline
\hline
 ($\langle x \rangle$, $\langle p \rangle$)   &   &  \crossmark  &    &  \checkmarkbis &   &  \checkmarkbis  \\
\hline
 ($\langle K \rangle$,$\langle V_H \rangle$)    &   &  \checkmarkbis &    &  \checkmarkbis &   &  \crossmark \\
 \hline
 ($\langle K_B \rangle$,$\langle Q_B \rangle$)   &   & \checkmarkbis &    & \checkmarkbis &   & \checkmarkbis \\
\hline
\end{tabular}
\caption{\label{table} The three scenarios D1, D2, D3 for the dynamics with and without disorder for $N=2$. Initial two lines: values of ($x_{01}$,$x_{02})$ and ($p_{01}$,$p_{02})$ for the initial non-equilibrium state. Last three lines: features of expectation values of ($\langle x \rangle$, $\langle p \rangle$), and of orthodox $(\langle K \rangle$, $\langle V_H \rangle$) and Bohmian ($\langle K_B \rangle$, $\langle Q_B \rangle$) energies, where symbols $\checkmarkbis$ and $\crossmark$ respectively indicate whether or not such expectation values allow one to identify $t_{eq}$.}
\end{table}

\subsubsection{Dynamics from initial position}
\label{D1}

\begin{figure*}
\centering
\begin{minipage}{1\linewidth}
\includegraphics[height=8cm,width=\linewidth]{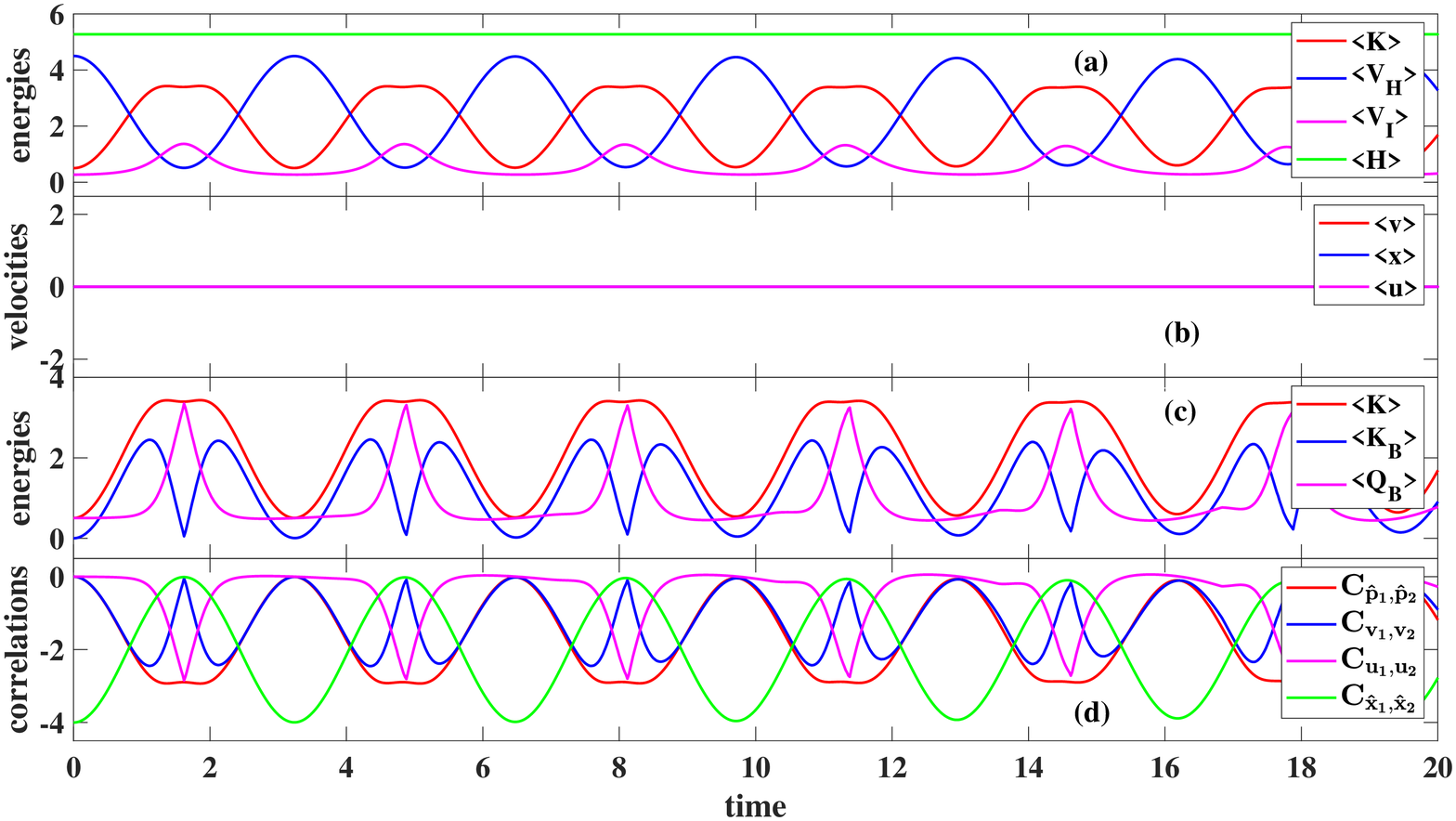}
\end{minipage}
\hfill
\begin{minipage}{1\linewidth}
\includegraphics[height=8cm,width=\linewidth]{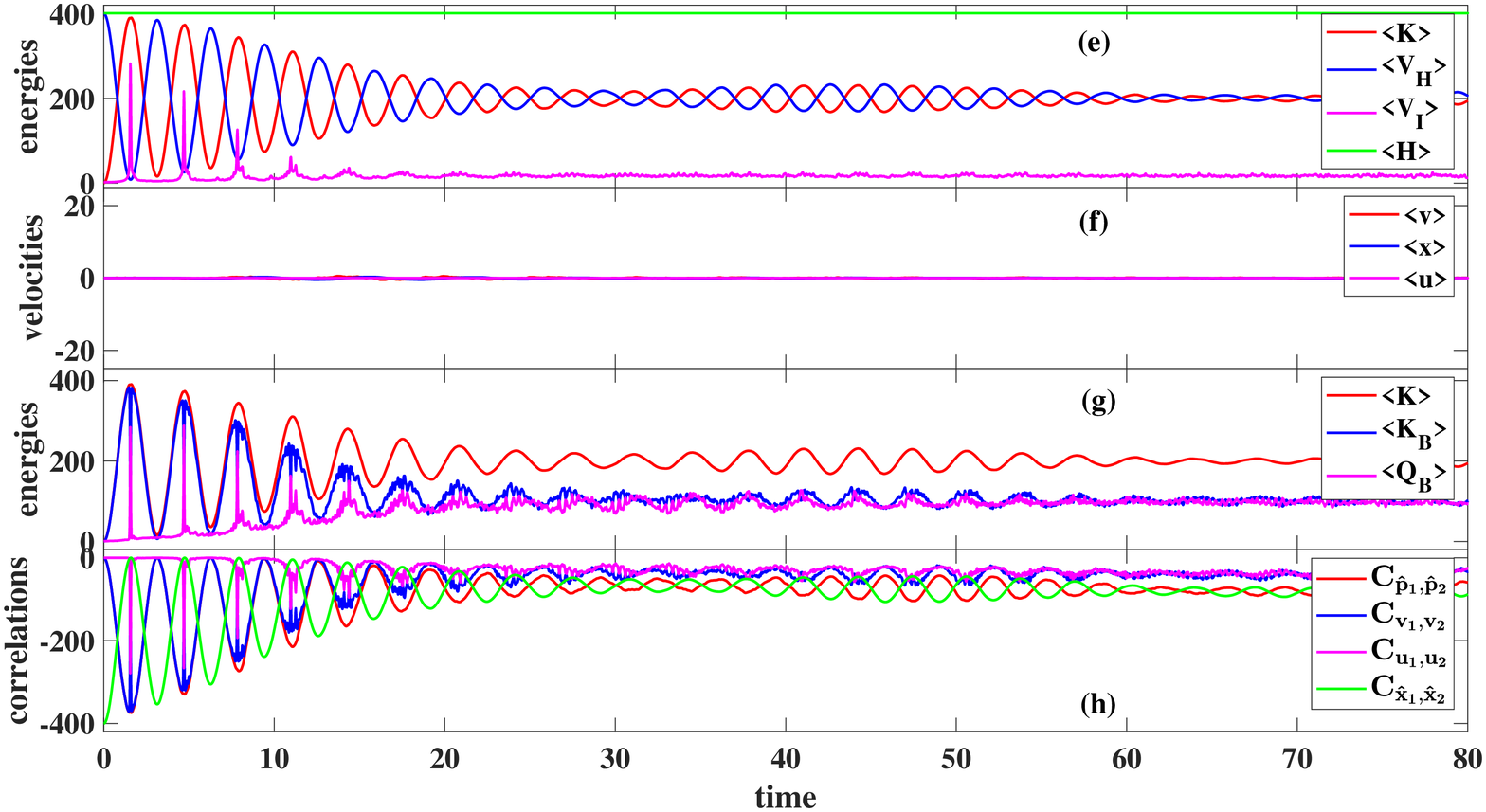}
\end{minipage}
\caption{Expectation values from the dynamics in the D1 scenario, with $(p_{01},p_{02})=(0,0)$. Top panels: few initial cycles under no disorder with a smaller $(x_{01},x_{02})=(-2,2)$; Bottom panels: full dynamics under disorder with a larger $(x_{01},x_{02})=(-20,20)$. Panels (a), (e) for energies: Orthodox kinetic energy $\langle K \rangle$, harmonic confining energy $\langle V_{H} \rangle$, Coulomb repulson energy $\langle V_{I} \rangle$, total energy $\langle H \rangle$ ($\langle V_{I} \rangle$ is magnified by $100$ in (e)). Panels (b), (f) for velocities: Bohmian velocity $\langle v \rangle$, position $\langle x \rangle$, osmotic velocity $\langle u \rangle$. Panels (c), (g) for kinetic energies comparison: Orthodox $\langle K \rangle$, Bohmian $\langle  K_B \rangle$, quantum potential $\langle Q_B \rangle$ energies. Panels (d), (h) for correlations: momentum $C_{\hat p_1,\hat p_2}$, Bohmian $C_{v_1,v_2}$ and osmotic $C_{u_1,u_2}$ velocities, position $C_{\hat x_1,\hat x_2}$.}
\label{fig12}
\end{figure*}

Let us first focus on the D1 dynamics without disorder in Fig. \ref{fig12}(a)-(d), which has $p_{01}=p_{02}=0$ and a small $x_{01} = -x_{02}=-2$. At $t=0$, \ref{fig12}(a) yields $\langle V_H \rangle (0) = x_{01}^2/2+x_{02}^2/2 + E_V/2=4.5$ and $\langle K \rangle (0) = p_{01}^2/2+p_{02}^2/2 + E_V/2=0.5$, where $E_V = \omega (n+1)$ is the ground state energy ($\omega=1$, $n=0$) which is equally shared among potential and kinetic terms. That is, the virial theorem in \eqref{vir2} is not satisfied (at any $t$) thanks to the non-equilibrium initial situation; however, the total energy in \eqref{vir1}, $\langle H \rangle (0) = 5.27$ since $\langle V_I \rangle (0)=0.27$, is conserved (at any $t$) thanks to the unitary evolution. In \ref{fig12}(c), at $t=0$, one has $\langle Q_B \rangle (0)=0.5$ and $\langle K_B \rangle (0)=0$ (since the initial velocities are zero, see \ref{fig12}(b)), so that indeed $\langle K \rangle (0)=0.5$ and \eqref{KQsum} is satisfied (at any $t$), while \eqref{KQhalf} does not apply here. Each D1-cycle has a $\pi$-period and three stages. For the first cycle: (i) at $t=0$ electrons are at $(x_1,x_2)=(-2,2)$, with minimum $\langle K \rangle$ and maximum $\langle V_{H} \rangle$; (ii) the dynamics pushes the electrons against each other until that, at $t=\pi/2$, they try binding together at $(x_1,x_2)=(0,0)$, which is avoided thanks to both exchange symmetry, Coulomb repulsion, and quantum potential, as indicated by the peak in $\langle V_I \rangle$ and in $\langle Q_B \rangle$, with maximum $\langle K \rangle$ and minimum $\langle V_{H} \rangle$ and $\langle K_B \rangle$; (iii) at $t= \pi$ electrons are back to $(x_1,x_2)=(-2,2)$ and a new cycle starts. Notice that $\langle K_B \rangle$ has a double peak around the peak of $\langle Q_B \rangle$ because the velocity acquires a first maximum from $t=0$ to $t=\pi/2$, which is the time electrons `stop' to reverse their movements, and a second maximum from $t=\pi/2$ to $t=\pi$. One sees how the quantum potential, acting `in-phase' with the Coulomb repulsion, carries the quantumness of the two-body entanglement. The correlations in \ref{fig12}(d) are a mirror of the above discussions; they are all negative since, as one variable increases, the other decreases in the D1-dynamics. In terms of moduli, at $t=0$, $C_{\hat x1,\hat x2}$ is at its maxima due to the maximum electron separation, and reaches its vanishing minima at $t=\pi/2$ when electrons are closest to each other. The three kinetic correlations are zero at $t=0$ thanks to the same initial zero velocity for the electrons; at $t=\pi/2$, since at that time it is only $\langle Q_B \rangle$ that contributes to $\langle K \rangle$, $C_{u1,u2}$ reaches its maximum while $C_{v1,v2}$ vanishes. Notice that \eqref{twoC} is satisfied at any $t$, while \eqref{Cb} does not apply here. In \ref{fig12}(b), $\langle v \rangle=\langle x \rangle=\langle u \rangle=0$ at any $t$ is a trivial consequence of the D1 dynamics being antidiagonal in the configuration space $x_1$$x_2$, with electrons initially equidistant, what anticipates that such terms reflect center of mass properties.

\begin{figure*}
\centering
\begin{minipage}{1\linewidth}
\includegraphics[height=8cm,width=\linewidth]{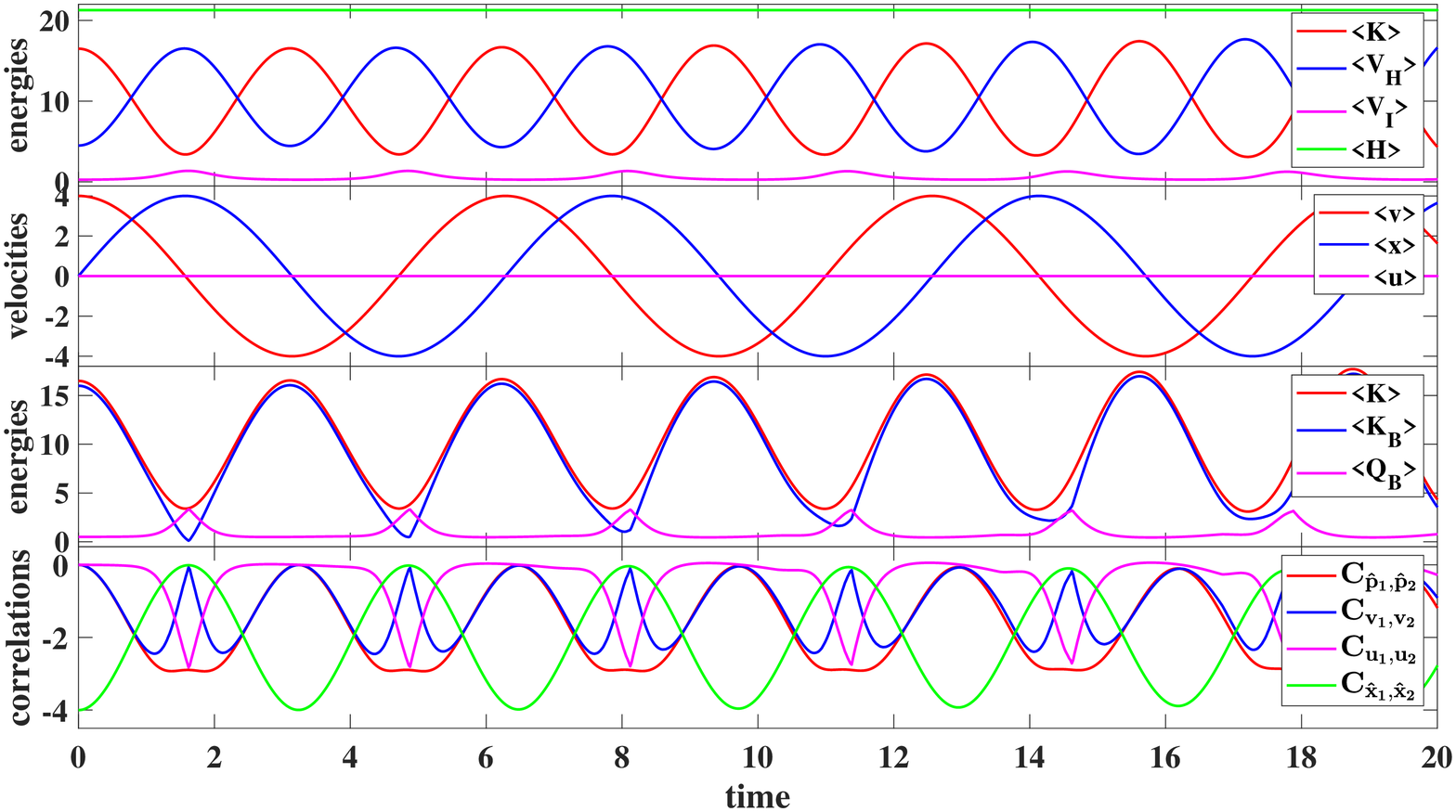}
\end{minipage}
\hfill
\begin{minipage}{1\linewidth}
\includegraphics[height=8cm,width=\linewidth]{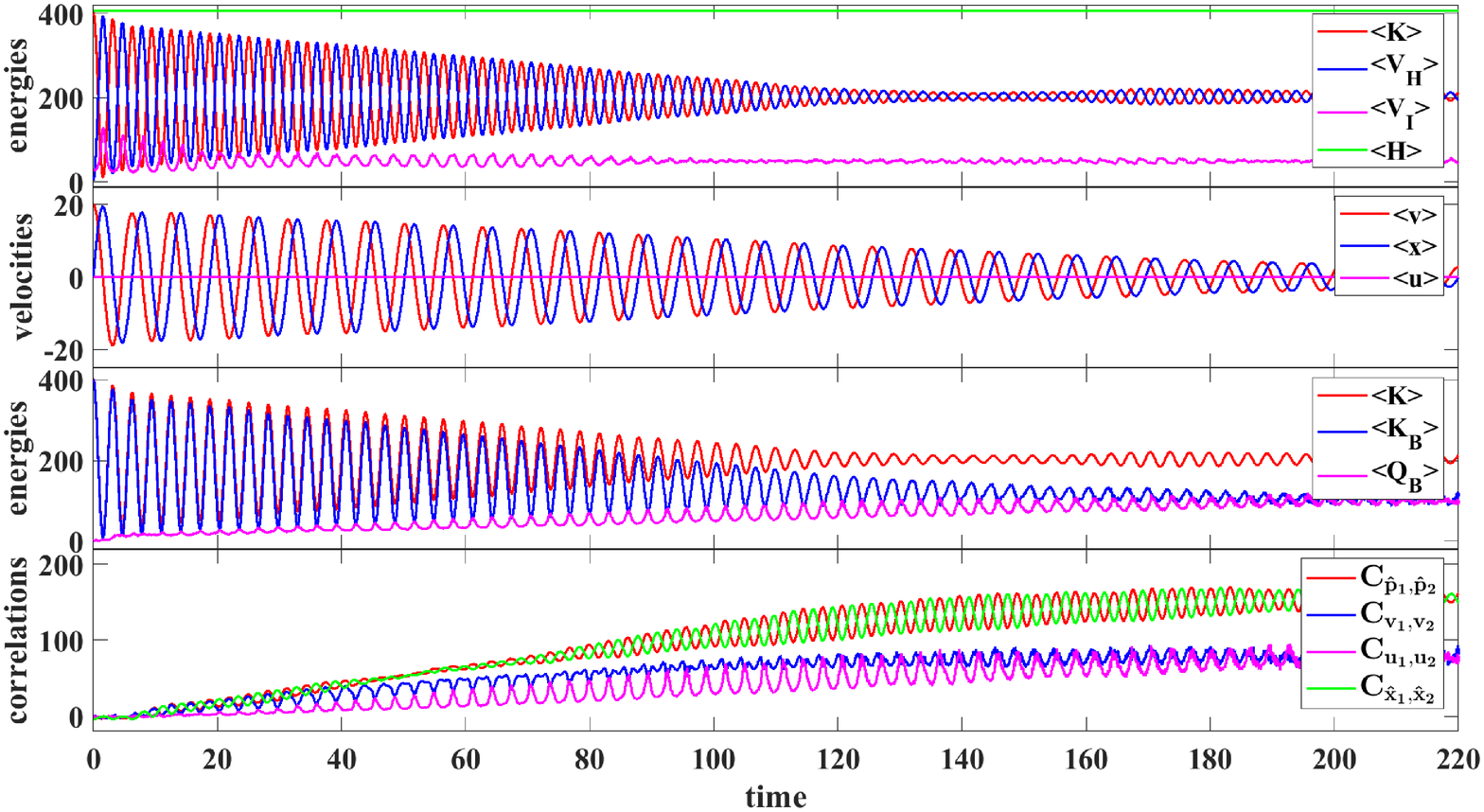}
\end{minipage}
\caption{Expectation values from the dynamics in the D2 scenario, with  $(x_{01},x_{02})=(-2,2)$. Top panels: few initial cycles under no disorder with a smaller $(p_{01},p_{02})=(4,4)$; Bottom panels: full dynamics under disorder with a larger $(p_{01},p_{02})=(20,20)$. Panels (a), (e) for energies: Orthodox kinetic energy $\langle K \rangle$, harmonic confining energy $\langle V_{H} \rangle$, Coulomb repulson energy $\langle V_{I} \rangle$, total energy $\langle H \rangle$ ($\langle V_{I} \rangle$ is magnified by $100$ in (e)). Panels (b), (f) for velocities: Bohmian velocity $\langle v \rangle$, position $\langle x \rangle$, osmotic velocity $\langle u \rangle$. Panels (c), (g) for kinetic energies comparison: Orthodox $\langle K \rangle$, Bohmian $\langle  K_B \rangle$, quantum potential $\langle Q_B \rangle$ energies. Panels (d), (h) for correlations: momentum $C_{\hat p_1, \hat p_2}$, Bohmian $C_{v_1,v_2}$ and osmotic $C_{u_1,u_2}$ velocities, position $C_{\hat x_1,\hat x_2}$.}
\label{fig34}
\end{figure*}

The full D1 dynamics with disorder is shown in Fig. \ref{fig12}(e)-(h). Thanks to the larger $x_{01}=-x_{02} =-20$ the oscillation amplitudes are larger but with the same $\pi$-period. The magnitude of $\langle V_{I} \rangle$ though remain similar to the no-disorder case, such that we magnify it by $100$ here, while the initial peaks in $\langle V_{I} \rangle$ and in $\langle Q_B \rangle$ as well as the minima in $\langle K_B \rangle$ become steeper in the disordered scenario. As the cycles succeed all expectation values overall reach quasi-stationary magnitudes once thermalization is set at $t_{eq} \approx 55$. In \ref{fig12}(e), while maintaining $\langle H \rangle$ constant as in \eqref{vir1}, kinetic and potential energies interchange their magnitudes until \eqref{vir2} becomes valid, and the virial theorem looks restated after thermalization, seeming an indication that a steady state is reached; the Coulomb repulsion $\langle V_{I} \rangle$ smears out and its peaks disappear after thermalization. In \ref{fig12}(g), while \eqref{KQsum} remains true at any $t$, the peaks in $\langle Q_B \rangle$ and minima in $\langle K_B \rangle$ also smear out and disappear after thermalization; most interestingly, this happens in a way as to satisfy \eqref{KQhalf}, allowing one to visualize the central result of our paper: in addition to the `restatement' of the virial theorem, thermalization also implies a kinetic energy equipartition, the hidden-variable signature of quantum thermalization. The trivial values of $\langle x \rangle=\langle v \rangle=\langle u \rangle \approx 0$ remain at any $t$ in \ref{fig12}(f), and so presents no feature to identify thermalization. The correlations in \ref{fig12}(h) are once more a mirror of the above discussions, so the maxima and minima of the initial cycles smear out after thermalization is set, when one also finds that, in addition to \eqref{twoC}, equation \eqref{Cb} is also satisfied, as well as $C_{\hat p_1,\hat p_2} \approx C_{\hat x_1,\hat x_2}$; the correlations remain negative, reaching small magnitudes after thermalization.
 
\subsubsection{Dynamics from initial velocity}
\label{D2}

\begin{figure*}
\centering
\begin{minipage}{1\linewidth}
\includegraphics[height=8cm,width=\linewidth]{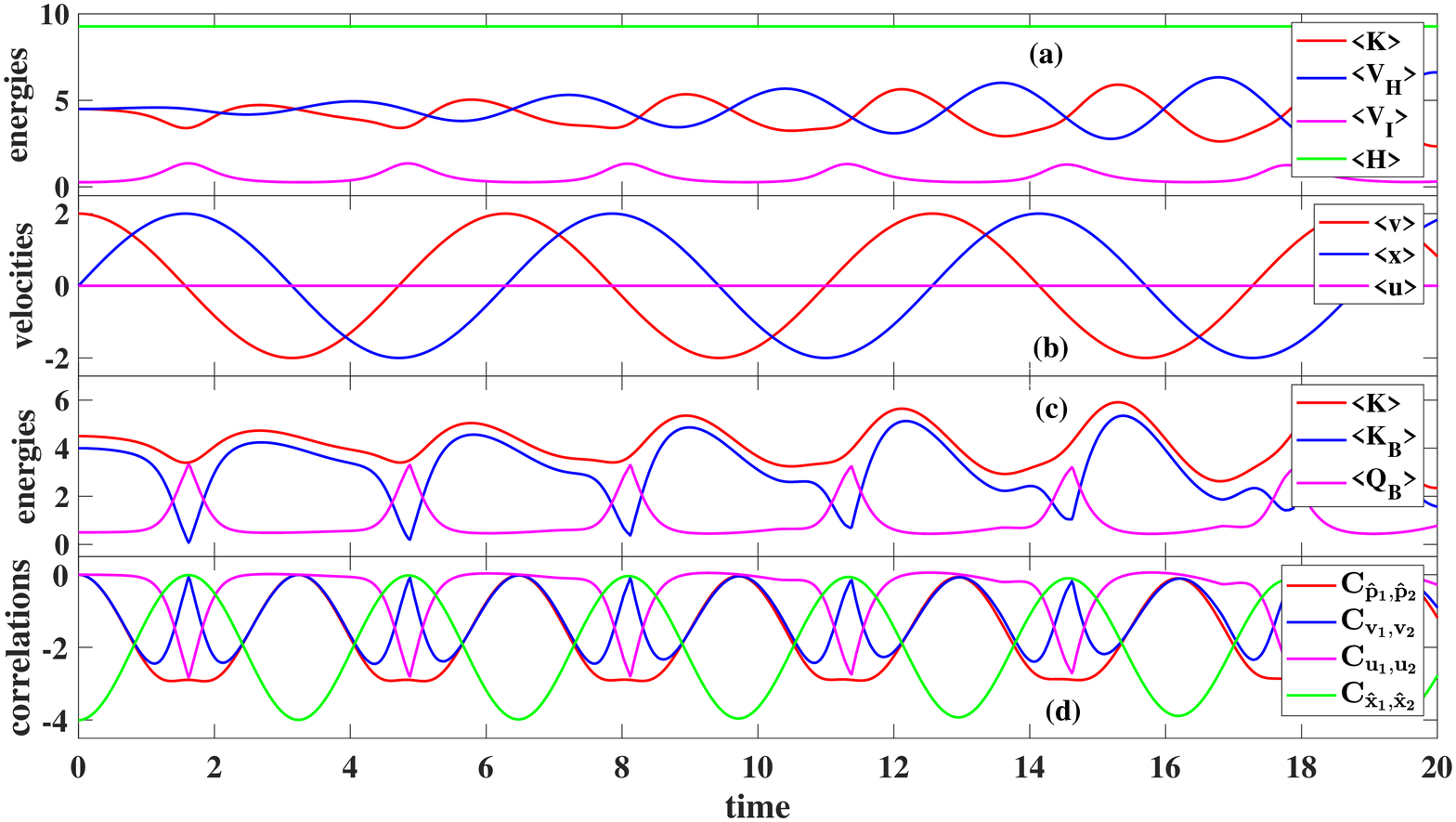}
\end{minipage}
\hfill
\begin{minipage}{1\linewidth}
\includegraphics[height=8cm,width=\linewidth]{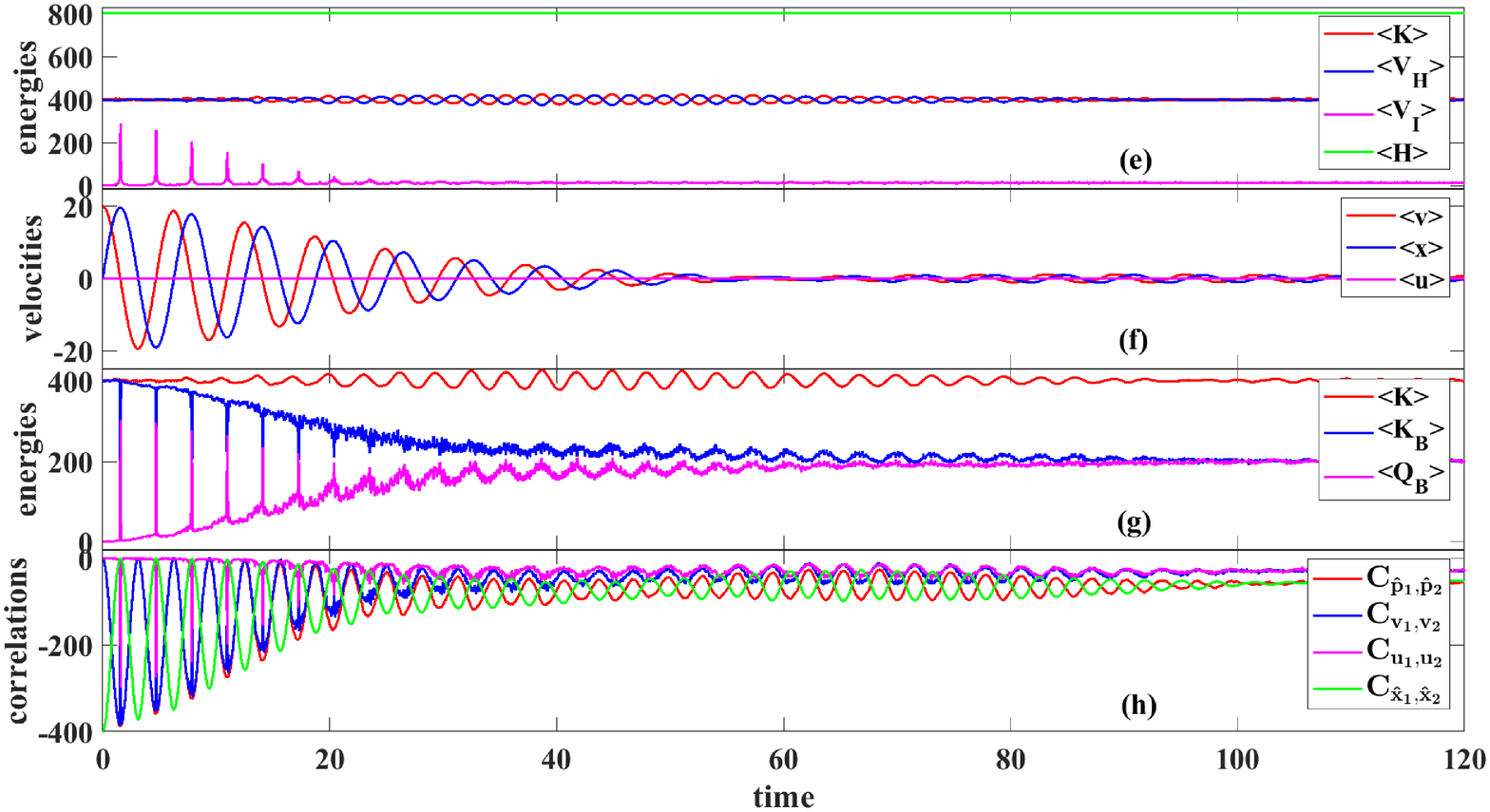}
\end{minipage}
\caption{Expectation values from the dynamics in the D3 scenario, with mixed $x_0$, $p_0$ influence. Top panels: few initial cycles under no disorder with small $(x_{01},x_{02})=(-2,2)$ and small $(p_{01},p_{02})=(2,2)$; Bottom panels: full dynamics under disorder with a large $(x_{01},x_{02})=(-20,20)$ and a large $(p_{01},p_{02})=(20,20)$. Panels (a), (e) for energies: Orthodox kinetic energy $\langle K \rangle$, harmonic confining energy $\langle V_{H} \rangle$, Coulomb repulson energy $\langle V_{I} \rangle$, total energy $\langle H \rangle$ ($\langle V_{I} \rangle$ is magnified by $100$ in (e)). Panels (b), (f) for velocities: Bohmian velocity $\langle v \rangle$, position $\langle x \rangle$, osmotic velocity $\langle u \rangle$. Panels (c), (g) for kinetic energies comparison: Orthodox $\langle K \rangle$, Bohmian $\langle  K_B \rangle$, quantum potential $\langle Q_B \rangle$ energies. Panels (d), (h) for correlations: momentum $C_{\hat p_1,\hat p_2}$, Bohmian $C_{v_1,v_2}$ and osmotic $C_{u_1,u_2}$ velocities, position $C_{\hat x_1,\hat x_2}$.}
\label{fig56}
\end{figure*}

Let us start the discussion of the D2 dynamics once more from the no-disorder case in Fig. \ref{fig34}(a)-(d), which has $x_{01}= -x_{02} =-2$ (as $(0,0)$ cannot be used in a fermionic trap) and a small $p_{01}=p_{02} = 4$ (such that both electrons start moving in the same direction). As such, at $t=0$, one still has $\langle V_H \rangle (0) = 4.5$ and $\langle V_I \rangle (0) = 0.27$, but a larger $\langle K \rangle (0) = 16.5$, built from $\langle K_B \rangle (0)=16$ and $\langle Q_B \rangle (0)=0.5$, so yielding $\langle H \rangle (0) = 21.27$. Each D2-cycle has a $2 \pi$-period and five stages. The first cycle is: (i) at $t=0$ electrons at $(x_1,x_2)=(-2,2)$ with minimum $\langle V_{H} \rangle$ and maximum $\langle K \rangle$; (ii) at $t=\pi/2$ electrons reach the positive turning point and try to collide at $(x_1,x_2)=(4,4)$, yielding peaks in $\langle V_{I} \rangle$ and in $\langle Q_B \rangle$, while $\langle K_B \rangle=0$ as the electrons stop at that time (the fact that the peak in $\langle Q_B \rangle$ is now at the minimum of $\langle K \rangle$ is the reason why a double peak is no longer seen in $\langle K_B \rangle$ contrarily to the D1-scenario); (iii) at $t=\pi$ electrons pass back at $(x_1,x_2)=(-2,2)$; (iv) at $t=3\pi/2$ the electrons reach the negative turning point and try to collide at $(x_1,x_2)=(-4,-4)$, with new peaks in $\langle V_{I} \rangle$ and in $\langle Q_B \rangle$ and again with $\langle K_B \rangle=0$; (v) at $t=2\pi$ the electrons are back to $(x_1,x_2)=(-2,2)$ and a new cycle starts. The cycle shows that, although the initial velocity is the same for both electrons, they acquire different velocities in the dynamics as one moves in favour and the other against the harmonic potential. Results in \ref{fig34}(b) are trivially understood from the previous analysis, confirming the center of mass character of those quantities: the D2 dynamics, being diagonal in the configuration space $x_1$$x_2$, has that $\langle x \rangle$ ranges from $\langle x \rangle=0$ when electrons are at $(2,-2)$ to $\langle x \rangle=4$ ($\langle x \rangle=-4$) when electrons are at the positive $(4,4)$ (negative $(-4,-4)$) turning point, while $\langle v \rangle$ is just out-of-phase and $\langle u \rangle=0$ from \eqref{int_u}. Interestingly the correlations in \ref{fig34}(d) do not change from the D1 dynamics. So let us explain the origin of the double-peak feature in $C_{v_1,v_2}$: at $t=0$, $C_{v_1,v_2}=0$ since both electrons have same initial velocity to the right; as time evolves, the left (right) electron gains (loses) velocity, so negatively increasing $C_{v_1,v_2}$, which reaches its first peak as the left electron crosses the origin; then the left electron also starts to decrease its velocity until both electrons reach the positive turning point, causing $C_{v_1,v_2}=0$ at $t=\pi/2$; electrons then start moving to the left, again with a higher velocity for the left electron, inducing a new negative increase in $C_{v_1,v_2}$, until the left electron passes by the origin at its highest velocity, causing the second peak in $C_{v_1,v_2}$.

The full D2 dynamics with disorder is seen in Fig. \ref{fig34}(e)-(h), in which the larger $p_{01}=p_{02}=20$ induces larger oscillation amplitudes with the same $2 \pi$-period, except for $\langle V_{I} \rangle$ which is once more magnified by $100$. Overall, the smearing out of the oscillations in $\langle K \rangle$ and $\langle V_{H} \rangle$, of the peaks in $\langle V_{I} \rangle$ and $\langle Q_B \rangle$, and of the minima in $\langle K_B \rangle$, happens as the system thermalizes, with such quantities seeming to reach stationary values at $t_{eq}\approx 220$. Thermalization in the D2 scenario also induces the feature $\langle x \rangle \approx \langle v \rangle \approx 0$ in \ref{fig34}(f) (compare with \ref{fig12}(f)). All four equations \eqref{KQsum}-\eqref{vir2} are also satisfied in the D2 scenario at $t > t_{eq}$, such that the virial theorem seems restated in \ref{fig34}(e), and the hidden-variable signature of thermalization is once more seen in \ref{fig34}(g). While \eqref{twoC} and \eqref{Cb} are also valid in the D2 scenario, as well as $C_{\hat p_1,\hat p_2} \approx C_{\hat x_1,\hat x_2}$, the correlations in \ref{fig34}(h) become positive (since both electrons move in the same direction) and stabilize at larger magnitudes (from the larger initial velocity).

\subsubsection{Dynamics from initial position and initial velocity}
\label{D3}

After detailing the scenarios D1 and D2 we will only describe here the distinct features of the D3 dynamics. Once more starting from the no-disorder case in Fig. \ref{fig56}(a)-(d), which has both a small $p_{01}=p_{02}=2$ and a small $x_{01} = -x_{02} =-2$. Such identical initial values obviously yield $\langle V_H \rangle (0) = \langle K \rangle (0) = 4.5$, and since $\langle V_I \rangle (0)=0.27$ one has $\langle H \rangle (0) = 9.27$, while $\langle K_B \rangle (0)=4$ and $\langle Q_B \rangle (0)=0.5$ are the values building $\langle K \rangle (0)$. The dynamics though is more involved since it is a superposition of the two previous scenarios, with cycles in a $2 \pi$-period. At the initial cycles, $\langle K \rangle$ looks almost constant since $\langle Q_B \rangle$ reaches magnitudes comparable to $\langle K_B \rangle$ (compare \ref{fig56}(c) with \ref{fig34}(c) and \ref{fig12}(c)), as $\langle K_B \rangle$ may still vanish whenever $\langle v \rangle=0$ in \ref{fig56}(b). The correlation plots in \ref{fig56}(d), interestingly, are identical to \ref{fig34}(d) and \ref{fig12}(d), while $\langle v \rangle$ and $\langle x \rangle$ in \ref{fig56}(b) oscillate in between the turning points at $(2,2)$ and $(-2,-2)$, with center-of-mass-like values.

The full D3 dynamics with disorder is shown in Fig. \ref{fig56}(e)-(h), which considers both a large $p_{01}=p_{02}=20$ and a large $x_{01}=-x_{02} = -20$. With such values the constancy of the kinetic and potential energies in \ref{fig56}(e) becomes evident such that, besides the trivial energy conservation in \eqref{vir1}, the virial theorem in \eqref{vir2} looks `as if' is always satisfied at any $t$, even before thermalization is set at $t_{eq}\approx 110$, mistakenly telling that some stationary state could be present since $t=0$. In other words, the onset of thermalization is no longer identifiable in the energy expectation values of \ref{fig56}(e) (where $\langle V_{I} \rangle$ is magnified again by $100$), although it remains identifiable in \ref{fig56}(f). The central result of our paper, however, is once more found in \ref{fig56}(g), where the kinetic energy equipartition in \eqref{KQhalf} is verified although following a different thermalizing path if compared to \ref{fig34}(g) and \ref{fig12}(g). The correlations in \ref{fig56}(h) are about the same as in the D1 scenario, where \eqref{Cb}, as well as $C_{\hat p_1,\hat p_2} \approx C_{\hat x_1,\hat x_2}$, apply once more.

\subsection{Revisiting the three scenarios within the center of mass framework}
\label{centerofmass}

\subsubsection{The issue of empirical indistinguishibility}
\label{issue}

All expectation values presented in the previous figures refer to a single particle but, thanks to the antisymmetry of $\Psi(\mathbf{x},t)$, there is no need to identify `which' particle. This is true no matter they are linked to Hermitian operators like $\hat K$ or $\hat p$ or to post-processings of weak values like $p_{W}(\mathbf{x},t)$, as the exchange symmetry of both phase and amplitude of $\Psi(\mathbf{x},t)$ is directly translated, respectively, to the current and osmotic velocities in \eqref{DEFvj}-\eqref{DEFuj} (and to the post-processing data). However, if one is interested on \textit{how} expectation values like $\langle v \rangle$ or $\langle u \rangle $  are obtained in a laboratory, one needs to differentiate between the ones computed from hermitian operators to the ones computed from post-processings of weak values, as indicated in Table \ref{table2}. 

\begin {table}
\label{tab:table}
\begin{tabular}{c|c|c|c}
\hline
 Expectation    & Eq.   &  Operator  &  From weak values \\
\hline 
$\langle p \rangle $  & \eqref{EVp}  &  $\hat p$ &  $p_{W}(\mathbf{x},t)$  \\
\hline
  $\langle K \rangle$    &  \eqref{OKE} &  $\hat K$ &   Re$[p_{W}(\mathbf{x},t)]^2$+Im$[p_{W}(\mathbf{x},t)]^2$  \\
\hline
\hline
  $\langle v \rangle $  &  \eqref{int_v}  &  $\crossmark$&  Re$[p_{W}(\mathbf{x},t)]$  \\
\hline
  $\langle u \rangle $  &  \eqref{int_u}  &  $\crossmark$ &  Im$[p_{W}(\mathbf{x},t)]$ \\
\hline
 $\langle K_B \rangle$   &  \eqref{BKE} & $\crossmark$ &   Re$[p_{W}(\mathbf{x},t)]^2$  \\
\hline
$\langle Q_B \rangle$   &  \eqref{BQP} & $\crossmark$ &   Im$[p_{W}(\mathbf{x},t)]^2$\\
\hline
\hline
\end{tabular}
\caption{\label{table2} The main expectation values relevant in the discussion of the kinetic energy equipartition, addressed in the previous figures. Expectation values marked with $\crossmark$ cannot directly be computed from an Hermitian operator, but they can be accessed by post-processing the local-in-position weak value of the momentum $p_{W}(\mathbf{x},t)$.}
\end{table}

\begin{figure*}
\centering
\begin{minipage}{1.0\linewidth}
\includegraphics[height=8cm,width=\linewidth]{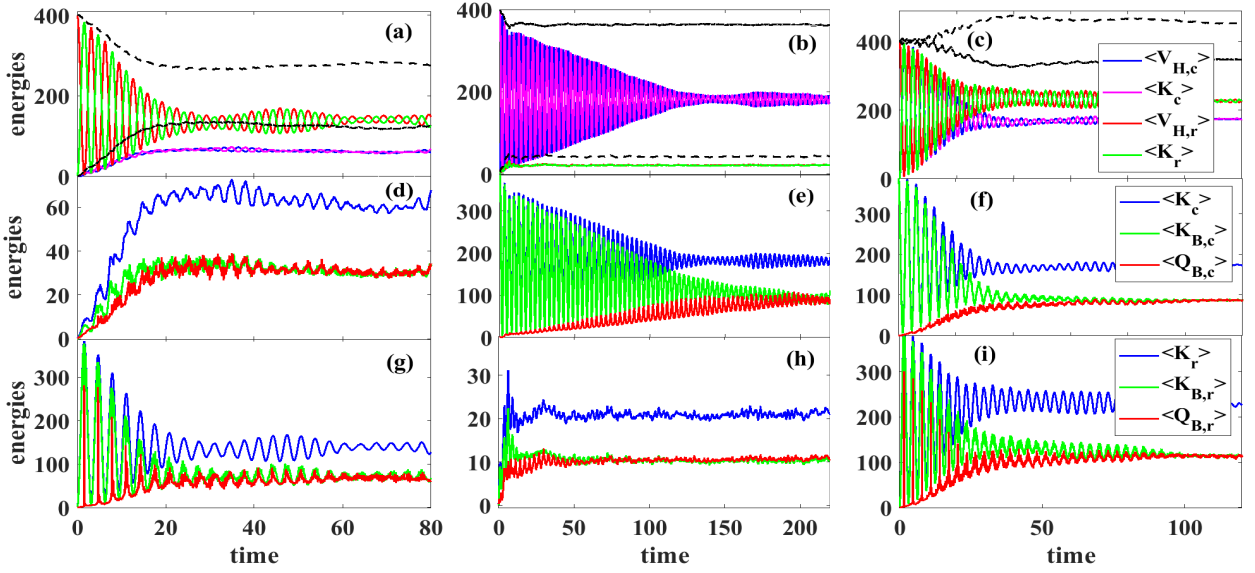}
\end{minipage}
\caption{Center of mass and relative energy expectation values for scenarios D1, D2, and D3, respectively, in left ((a),(d),(g)), middle ((b),(e),(h)), and right ((c),(f),(i)) panels. Upper panels: kinetic energy $\langle K_{c}\rangle$ and $\langle K_{r} \rangle$, confining potential energy $\langle V_{H,c}\rangle$ and $\langle V_{H,r}\rangle$), respective total energy $E_c=\langle K_c \rangle + \langle V_{H,c} \rangle$ and $E_r=\langle K_r \rangle + \langle V_{H,r} \rangle$ (black lines); $\langle V_{I,r} \rangle$ is not included. Middle panels: Bohmian kinetic $\langle K_{B,c} \rangle$ and quantum potential $\langle Q_{B,c} \rangle$ energies in comparison with orthodox $\langle K_{c}\rangle$ center of mass energies. Lower panels: Bohmian kinetic $\langle K_{B,r} \rangle$ and quantum potential $\langle Q_{B,r} \rangle$ energies in comparison with orthodox $\langle K_{r}\rangle$ relative energies. Legend in (c) applies to (a),(b); legend in (f) applies to (d),(e); legend in (i) applies to (g),(h).} 
\label{fig7}
\end{figure*}

Let us ilustrate the problem with $N=2$. The experimental weak value $p_{W,1}(x_1,x_2,t)$ requires two measurements: a first weak measurement linked to $\hat p_1$, plus a second strong measurement linked to $\hat x_1$ and $\hat x_2$. Such a weak value is obtained after a pre-selection (repeating the same experiment for a large number of identically prepared initial wave functions) plus a post-selection (to obtain an ensemble value of all weakly measured momentum whose subsequent strongly measured position yields a specific location); a similar procedure is required for getting $p_{W,2}(x_1,x_2,t)$ by interchanging the measurements on particles $1$ and $2$. Although $\langle p_{W,1} \rangle=\int dx_1\;dx_2  p_{W,1}(x_1,x_2,t) |\Psi(x_1,x_2,t)|^2$ is the same as $\langle p_{W,2} \rangle = \int dx_2\;dx_1  p_{W,2}(x_2,x_1,t) |\Psi(x_2,x_1,t)|^2$, their empirical evaluation requires one to identify which are the particles $1$ and $2$ being weakly or strongly measured, which is impossible for systems with indistinguishable particles. One should keep in mind though that while Bohmian particles are ontologically distinguishable, the Bohmian dynamical laws ensure that expectations values are empirically indistinguishable. 

To handle this issue we consider an effective single-particle weak value, $\tilde p_{W,j}(x_k,t)$, by defining $\mathbf{X_k} = (x_1,..,x_{k-1},x_{k+1},..,x_N)$ such that $\mathbf{x}=(x_k,\mathbf{X_k})$, as
\begin{eqnarray}
\tilde p_{W,j}(x_k,t) &=& \frac{\int  d\mathbf{X_k} \; p_{W,j}(x_k,\mathbf{X_k},t) |\Psi(x_k,\mathbf{X_k},t)|^2}{P(x_k,t)}, \label{wvpc} \\
 P(x_k,t)                    &=& \int  d\mathbf{X_k} \; |\Psi(x_k,\mathbf{X_k},t)|^2. \label{wfpc}
\end{eqnarray}
 It is straightforward to show that
\begin{eqnarray}
\langle \tilde p_{W,j} \rangle &=& \int dx_k\;  \tilde p_{W,j}(x_k,t) P(x_k,t) \rangle \nonumber\\
			               &=& \int dx_k d\mathbf{X_k} \;p_{W,j}(x_k,\mathbf{X_k},t) |\Psi(x_k,\mathbf{X_k},t)|^2 \nonumber \\
                                               &=& \langle p_{W,j} \rangle.
\end{eqnarray}
Here we only need the case $j=k$; the general case $j \ne k$ is found elsewhere \cite{ourPRA}. We remind that $\langle \tilde p_{W,k} \rangle=\langle p_{W,k} \rangle$ does not imply that $\tilde p_{W,k}(x_k,t)$ can promptly be used in place of $p_{W,k}(x_k,\mathbf{X_k},t)$ to obtain expectation values like $\langle v \rangle$ or $\langle u \rangle $. This is only the case when the degree $x_k$ is decoupled from the rest $N-1$ degrees in $\mathbf{X_k}$, that is, when one has $\Psi(x_k,\mathbf{X_k},t) = \psi_k(x_k,t)\psi_\mathbf{k}(\mathbf{X_k},t)$ . Then, it is straightforward to show that $p_{W,k}(x_k,\mathbf{X_k},t)=\tilde p_{W,k}(x_k,t)$ with $P(x_k,t)=|\psi_k(x_k,t)|^2$, so that the evaluation of the weak value linked to the weak momentum and position of the particle $x_k$ does not require the measurements of the others positions $\mathbf{X_k}$, but, still requires to identify the particle $x_k$. This last difficulty can be solved by dealing with the center of mass frame. 

It is well-know \cite{BMxav,BMthomas,BMthomas2,BM,ana1,ana2} that our trap Hamiltonian in the absence of the disorder potential $V_D (\mathbf{x})$, that is, when $H (\mathbf{x}) \equiv K(\mathbf{x}) + V_H (\mathbf{x}) + V_I (\mathbf{x})$, can be written in terms of center of mass $x_c$ and relative  $\mathbf{X_r}=(x_{r,1},x_{r,2},..,x_{r,N-1})$ coordinates, where $x_{c} = \sum_{i=1}^{N} x_i / N$ and $x_{r,j} = x_j-x_{j+1}$. In such a situation (top panels in all previous figures), the discussion above promptly applies by taking $x_c \equiv x_k$ and $\mathbf{X_r} \equiv \mathbf{X_k}$. Then one indeed could consider $\tilde  p_{W,c}(x_c,t)$, which one has experimental access, instead of the intricate $ p_{W,c}(x_c,\mathbf{X_r},t)$, to compute quantities from Table \ref{table2}
like $\tilde v_c^2(x_c,t)=\text{Re}(\tilde p_{W,c}(x_c,t))^2$ and $\tilde u_c^2(x_c,t)=\text{Im}(\tilde p_{W,c}(x_c,t))^2$, instead of $v_c^2(x_c,\mathbf{X_r},t)$ or $u_c^2(x_c,\mathbf{X_r},t)$. One would also have $H = H_c + H_\mathbf{r}$ and $\Psi(x_c,\mathbf{X_r},t) = \psi_c(x_c,t)\psi_\mathbf{r}(\mathbf{X_r},t)$,  such that $\tilde  p_{W,c}(x_c,t) =p_{W,c}(x_c,\mathbf{X_r},t)$ in \eqref{wvpc} and $P(x_c,t) = |\psi_c(x_c,t)|^2$ in \eqref{wfpc}.  That is, a weak measurement of the momentum of the center of mass followed by a strong measurement of its position, without measuring the other $N-1$ degrees, should yield information about the $N$-body trap. The kinetic energy equipartition discussed so far should remain valid in any frame of reference. So the question one needs to pose now is whether or not the inclusion of disorder, which drives the system towards thermalization (bottom panels in all previous figures), induces a coupling between center of mass and relative coordinates. 

\subsubsection{Results in the center of mass frame}
\label{comresults}

To answer that question we once more make use of $N=2$, with $\mathbf{X_r}=(x_{r,1}) \equiv x_r$. The disordered trap Hamiltonian becomes simply
\begin{equation}
H(x_c,x_r)=H_c(x_c)+H_r(x_r)+V_D(x_c,x_r),
\label{Hcomrel}
\end{equation}
with 
\begin{eqnarray}
H_c(x_c) &=&- \frac{1}{4}\frac{\partial^{2}}{\partial x_c^{2}} +\omega^{2} x_c^{2}, \label{hcom} \\
H_r(x_r) &=& - \frac{\partial^{2}}{\partial x_r^{2}} + \frac{1}{4}\omega^{2}x_r^{2} + \frac{1}{|x_r+\alpha|}; \label{hrel}
\end{eqnarray}
 With $V_D(x_c,x_r)=0$, one would obtain $\Psi(x_c,x_r,t) = \psi_c(x_c,t) \psi_r(x_r,t)$ with $i \partial \psi_c(x_c,t)/\partial t =H_c \psi_c(x_c,t)$ and $i \partial \psi_r(x_r,t)/\partial t = H_r \psi_r(x_r,t) $.  Figure \ref{fig7} shows the dynamics for the scenarios D1, D2, D3, respectively, at left, middle, right panels, in which $\langle V_{H,c} \rangle$ and $\langle K_c \rangle$ ($\langle V_{H,r} \rangle$ and $\langle K_r \rangle$) label the confining potential and kinetic components of \eqref{hcom} (of \eqref{hrel}); the component related to the Coulomb term $\langle V_{I,r} \rangle$ in \eqref{hrel} is not shown since, as we have seen in all previous figures, it remains small. The labels $\langle K_{B,c}\rangle =\langle (v_1+v_2)^2 \rangle / 4$ and $\langle K_{B,r} \rangle =\langle (v_1-v_2)^2 \rangle / 4$, as well as $\langle Q_{B,c} \rangle =\langle (u_1+u_2)^2 \rangle / 4$ and $\langle Q_{B,r}\rangle =\langle (u_1-u_2)^2\rangle / 4$, respectively for the center of mass and relative components of both Bohmian kinetic and quantum potential energies, are also employed. 

The upper panels (Fig. \ref{fig7}-(a),(b),(c)) show that, indeed, both center of mass and relative coordinates thermalize, that is, $\langle V_{H,r} \rangle $ and $\langle K_r \rangle$ on one hand, while $\langle V_{H,c} \rangle$ and $\langle K_c \rangle$ on the other hand, reach about same stationary values at the same $t_{eq}$, with relative energies being higher in scenarios D1 and D3; the opposite happens in the D2 dynamics, where the relative energies also present a much smaller $t_{eq}$. The black lines, which present the total energies $E_c=\langle K_c \rangle+ \langle V_{H,c} \rangle$ and $E_r=\langle K_r \rangle + \langle V_{H,r} \rangle$, show that, in fact, center of mass and relative coordinates are \textit{not} decoupled at $t<t_{eq}$, since their respective energies do not remain constant in the time evolution. So, in principle, the picture made in the previous subsection should not apply. However, such energies do stabilize after $t_{eq}$, seemingly indicating that thermalization implies a negligible dependence of $x_c$ on $x_r$. This should not come as a surprise since, first, the physics should not be different by changing a frame of reference; second, as the disorder potential is random and does not privilege any degree of freedom, it seems natural to expect that, while coupling is to be found at the beginning, the time evolution should homogeneously spread the probabilities over the configuration space, no matter if among $x_1$ and $x_2$ or $x_c$ and $x_r$. So, the empirical measurements as pictured in the previous subsection do apply at thermalization and, more importantly, the negligible dependence of $x_c$ on the relative coordinates becomes an even more robust result for thermalized systems with large $N$  \cite{clas}. The middle (Fig. \ref{fig7}-(d),(e),(f)) and lower (Fig. \ref{fig7}-(g),(h),(i)) panels, respectively, confirm that the kinetic energy equipartition remains verified for the center of mass and relative coordinates, for each of the three scenarios, and it should also be verified for any $N$ in this new reference frame. One also has to remind, though, that thermalization has been studied in small systems with as little as 6 \cite{1DBosegases3}, 5 \cite{therm_rigol}, or 2-4 bosons \cite{lea_offdiag,lea_howmany}, and even single-particle systems \cite{singleparticle,thermalization_singleparticle,preprint}.

\section{Conclusions}
\label{conclusion}

Weak values have gradually been transitioning from a theoretical curiosity to a practical tool in the laboratory allowing novel characterizations of quantum systems, as they can provide information beyond the traditional expectation values linked to Hermitian operators. In particular, we have shown that  weak values of the momentum post-selected in positions, without linking the discussion to any specific ontology but re-using the mathematical machinery of the Bohmian and Stochastic quantum theories,  can be used as a relevant tool to characterize quantum thermalization in closed systems. As example, we have addressed the monopole oscillations in the configuration space of a two-electron harmonic trap under random disorder, under different initial conditions employed to initiate three distinct nonequilibrium dynamics. 

The expectation values from the orthodox operators not always can be employed to access the onset of thermalization. For example, $\langle v \rangle$ and $\langle x \rangle$ in Fig. \ref{fig12}(f) for the D1 scenario, or $\langle K \rangle$ and $\langle V_H \rangle$ in Fig. \ref{fig56}(e) for the D3 scenario; on the other hand, for the D2 scenario, both Figs. \ref{fig34}(e)-(f) identify such onset. The differences among these dynamics and their path to thermalization result from the different initial conditions, in which in the configuration space either the dynamics is antidiagonal (D1), diagonal (D2), or both (D3). 

On the other hand, the onset of thermalization is always accessible from its hidden-variable signature in every scenario, irrespective to initial conditions. Not only \eqref{KQsum} is obviously satisfied at any time, but the kinetic energy equipartition in \eqref{KQhalf}, stating that Bohmian kinetic and quantum potential energies should become equal, with each equalizing half of the orthodox kinetic energy, is always true after thermalization is set. This is the same of saying that the squared values of osmotic and current velocities become equal, with each equalizing half of the squared momentum, which also implies that the correlations obey \eqref{Cb} in addition to \eqref{twoC}; the validity of \eqref{vir2} in addition to \eqref{vir1} can be taken as a restatement of the virial theorem when reaching some steady state.

These hidden variables, linked separately to amplitude (osmotic) and phase (current) components of the many-body wave function, are not linked to orthodox operators, but are accessible in the laboratory through a post-proccessing of local-in-position weak values protocols for momentum and kinetic energy; the real and imaginary parts of the momentum weak value are respectively tied to the current and osmotic hidden variables. Thermalization is, so to speak, a manifestation of both real (amplitude) and imaginary (phase) parts of the wavefunction becoming completely and basically homogeneously spread through the whole configuration space, making it hard for one to differentiate among one or another. 

In order to properly understand the merits of our work it is essential to notice that all hidden-variable results in Secs. \ref{theory} and \ref{equipartition} are not  just visualizing tools, but they open a new unexplored link between theoretical predictions and empirical data. Notice that the use of term `hidden variables' has no ontological implication (only historical reasons). All quantum theories with empirical agreement with experiments do exactly predict the same weak values. In simpler words, the link between theoretical predictions and empirical data do not need to choose any particular ontology. To emphasize the accessibility of weak values in the laboratory, we  have also addressed, by moving to the center of mass frame of reference, how the weak values of such center of mass can be employed to approach larger systems with $N$ identical particles, where the novel kinetic energy equipartion here presented should also be verified.

\begin{acknowledgments}
This research was funded by Spain's Ministerio de Ciencia, Innovaci\'on y Universidades under Grant  PID2021-127840NB-I00 (MICINN/AEI/FEDER, UE), Generalitat de Catalunya and FEDER for project 001-P-001644 (QUANTUMCAT), European Union's Horizon 2020 research and innovation programme under Grant 881603 GrapheneCore3.
\end{acknowledgments}



\begin{thebibliography}{99}
\bibitem{param} All figures in the paper consider $\omega=1.0$, $\alpha=0.01$, $\sigma_{1}=\sigma_{2}=1/\sqrt{\omega}=1.0$, $\gamma_D=25$, $\sigma_D \approx \Delta x$, with $\Delta x$ being the position resolution of the grid, defined in a way that its momentum resolution is $\Delta p \approx \Delta x$. The initial wave packet is antisymmetrized, such that the label $j$ in quantities like $\langle x_{1,2} \rangle$ is redundant. The final simulation time is $t_f = 100$ for the D1 scenario, $t_f = 250$ for the D2 scenario, and $t_f = 150$ for the D3 scenario, with time step $\Delta t=0.001$. Simulation boxes are $[-60,60]$ with $2048$ points for each degree of freedom in the configuration space $x_1$$x_2$, so that for $N=2$ its size is $\approx 4 \; 10^6$.
\end{thebibliography}
\end{document}